\documentclass[aps,preprint,nofootinbib,preprintnumbers,eqsecnum]{revtex4-1}

\usepackage[usenames, dvipsnames]{color}

\usepackage[
	  pagebackref=false,
	  colorlinks=true,
      linkcolor=blue,
      urlcolor=blue,
      filecolor=black,
      citecolor=red,
      pdfstartview=FitV,
      pdftitle={},
        pdfauthor={},
        pdfsubject={},
        pdfkeywords={},
        pdfpagemode=None,
        bookmarksopen=true
      ]{hyperref}

\usepackage[normalem]{ulem}
\usepackage{amsmath}
\usepackage{enumerate}
\usepackage{amsfonts}
\usepackage{epsfig}
\usepackage{mathbbol}

\def\Tr{\mathop{\rm Tr}}
\def\tr{\mathop{\rm tr}}

\newcommand\ket[1]{\ensuremath{\lvert{#1}\rangle}}
\newcommand\bra[1]{\ensuremath{\langle{#1}\rvert}}

\newcommand{\be}{\begin{equation}}
\newcommand{\ee}{\end{equation}}
\newcommand{\bea}{\begin{eqnarray}}
\newcommand{\eea}{\end{eqnarray}}
\newcommand{\bega}{\begin{gather}}
\newcommand{\eega}{\end{gather}}

\newcommand{\bi}{\begin{itemize}}
\newcommand{\ei}{\end{itemize}}
\newcommand{\ben}{\begin{enumerate}}
\newcommand{\een}{\end{enumerate}}
\newcommand{\bca}{\begin{cases}}
\newcommand{\eca}{\end{cases}}
\newcommand{\bln}{\begin{align}}
\newcommand{\eln}{\end{align}}
\newcommand{\bst}{\begin{split}}
\newcommand{\est}{\end{split}}
\def\ie{\begin{equation}\begin{aligned}}
\def\fe{\end{aligned}\end{equation}}
\newcommand{\bma}{\le(\begin{matrix}}
\newcommand{\ema}{\end{matrix}\ri)}
\newcommand{\bwt}{\begin{widetext}}
\newcommand{\ewt}{\end{widetext}}

\newcommand\al{{\alpha}}
\def\b{{\beta}}

\newcommand\de{{\ensuremath{{\delta}}}}

\newcommand\da{{\dagger}}

\newcommand\ov{\over}

\def\le{\left}
\def\ri{\right}

\newcommand\sA{{\ensuremath{{\mathcal A}}}}
\newcommand\sB{{\ensuremath{{\mathcal B}}}}

\newcommand\sH{{\ensuremath{{\mathcal H}}}}

\newcommand\sO{{\ensuremath{{\mathcal O}}}}
\newcommand\sP{{\ensuremath{{\mathcal P}}}}
\newcommand\sQ{{\mathcal Q}}

\newcommand\sS{{\mathcal S}}

\usepackage{tensor}
\usepackage{braket}
\usepackage{comment}

\newcommand{\bid}{{\mathbf 1}}

\newcommand{\seq}{{s_{\rm eq}}}

\def\XXint#1#2#3{{\setbox0=\hbox{$#1{#2#3}{\int}$}
     \vcenter{\hbox{$#2#3$}}\kern-.5\wd0}}

\begin{document}

\title{A dynamical mechanism for the Page curve from quantum chaos 
}

\preprint{MIT-CTP/5175}

\author{Hong Liu and Shreya Vardhan}
\affiliation{Center for Theoretical Physics, \\
Massachusetts
Institute of Technology,
Cambridge, MA 02139 }

\begin{abstract}
 \noindent 
If the evaporation of a black hole formed from a pure state is unitary, the entanglement entropy of the Hawking radiation 
should follow the Page curve, increasing from zero until near the halfway point of the evaporation, 
and then decreasing back to zero. The general argument for the Page curve is based on the assumption that the quantum state 
of the black hole plus radiation during the evaporation process is  typical. In this paper, we show that the Page curve can result from a simple dynamical input in the evolution of the black hole, based on a recently proposed signature of quantum chaos,
without resorting to typicality. Our argument is based on what we refer to as the ``operator gas'' approach, 
which allows one to understand the evolution of the microstate of the black hole from generic features of the Heisenberg
evolution of operators. One key feature which leads to the Page curve is the possibility of dynamical processes 
where operators in the ``gas'' can ``jump'' outside the black hole, which we refer to as void formation processes. 
Such processes are initially exponentially suppressed, but dominate after a certain time scale, which can be 
used as a dynamical definition of the Page time. In the Hayden-Preskill protocol for young and old black holes, we show that 
void formation is also responsible for the transfer of information from the black hole to the radiation. 
 We conjecture that void formation may  provide  a microscopic explanation for the recent semi-classical prescription of including islands in the calculation of the entanglement entropy of the radiation.

\end{abstract}

\today

\maketitle

\tableofcontents

\section{Introduction}
\begin{figure}[!h]
\begin{center}
\includegraphics[width=6cm]{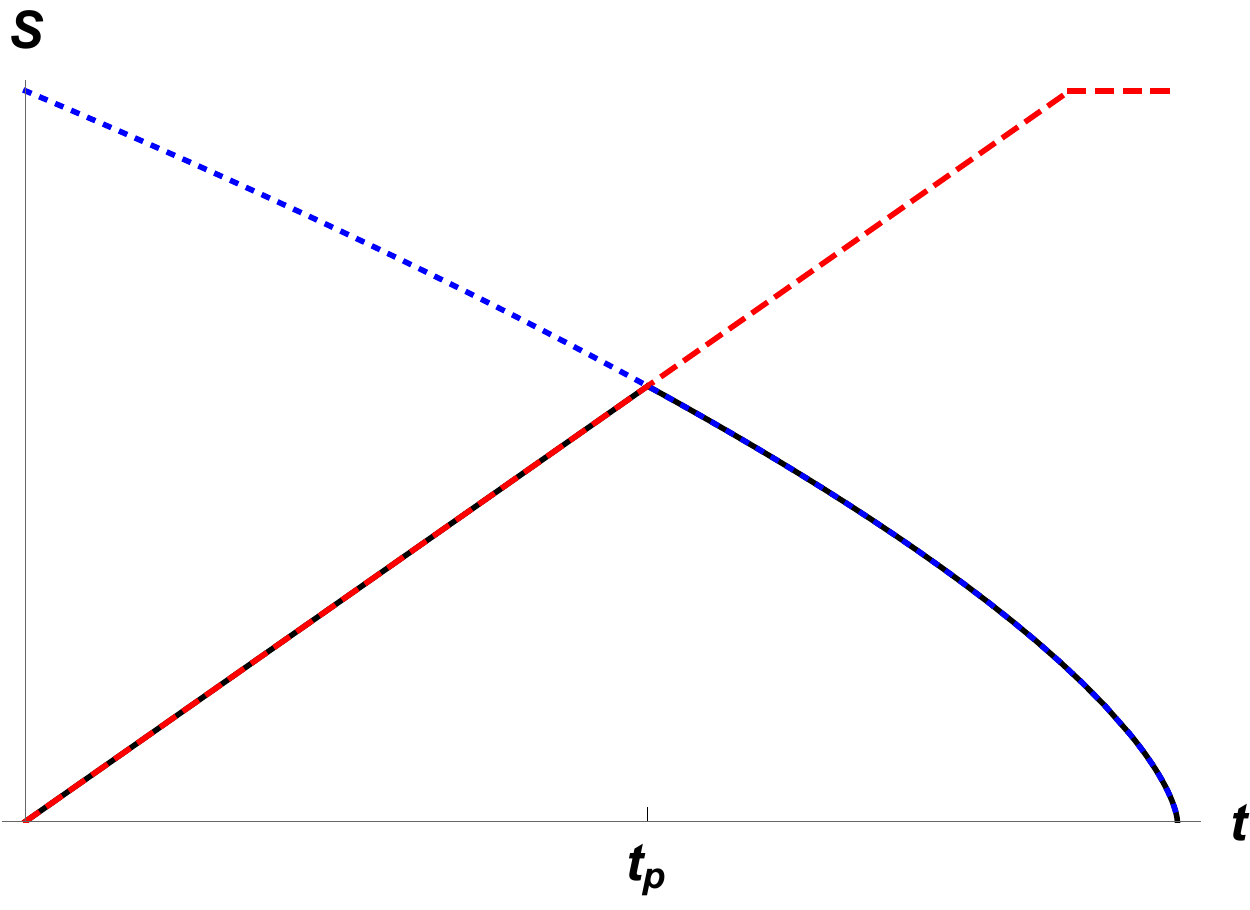}
\caption{The Page curves of the black hole and the radiation. At $t=0$, the black hole consists of the whole system and is in a pure state. The dotted line is the semi-classical entropy of the black hole from its horizon area. The dashed line is the entropy of the semi-classical radiation from Hawking's calculation. The solid curve is the entanglement entropy for the black hole and the radiation in a full quantum description. It should be seen as two curves, one for the black hole and one for the radiation, which coincide as required by unitarity.  The Page time $t_p$ refers to time scale where the solid curve  turns around from increasing to decreasing with time.  
}
\label{fig:page}
\end{center}
\end{figure}

Stephen Hawking's observation that a black hole emits thermal radiation leads to an apparent contradiction with the unitarity of quantum mechanics~\cite{Hawking, Hawking2}: when a black hole formed by the gravitational collapse of a pure state evaporates completely, one is left with only the emitted thermal radiation, which appears to be in a mixed state. 
For the final state to be pure, there must be global quantum correlations among different parts of the radiation which are not visible in Hawking's semi-classical calculations.
 Later, Don Page~\cite{page} pointed out a further consequence of the unitarity of an evaporation process:
the entanglement entropy of the radiation will initially increase as predicted by Hawking, but will have to decrease near the halfway point of the evaporation process, 
following what is now widely referred to as the Page curve. See Fig.~\ref{fig:page}. The time scale at the turning point is referred to as the Page time. 

The argument for the Page curve is very simple. Consider a quantum system $L = B \cup R$ 
with the Hilbert space $\sH_L = \sH_{B} \otimes \sH_{R}$, with the dimensions of $\sH_{B} $ and $\sH_{R} $ respectively equal to $d_{B}$ and $d_{R}$. Then on averaging the von Neumann entropies $S_B, S_R$ for the $B$ and $R$ subsystems over all pure states of $L$ with the Haar measure, one finds\footnote{The expression below
is the leading approximation in the regime $d_B \ll d_R$ or $d_R \ll d_B$.}~\cite{page2,lubkin,Lloyd}
\be \label{avEn}
\overline{S_B} = \overline{S_R} = {\rm min} (\sS_B, \sS_R) + \cdots , \qquad \sS_B = \log d_B , \quad \sS_R = \log d_R 
\ee
where $\sS_{B,R}$ are the ``coarse-grained entropies'' of the $B$ and $R$ subsystems. 
For an evaporating black hole, one takes $B$ and $R$ to be the black hole and radiation subsystems respectively. {The Hilbert space of both $B$ and $R$} changes with time. At $t=0$, $B (t=0) = L$ while $R(t=0)$ is empty, and as time goes on the degrees of freedom in $B(t)$ slowly go over to $R(t)$, until the black hole has completely evaporated. 
The Page curve then follows from~\eqref{avEn} if one assumes that during the evaporation process, the state of the black hole plus radiation at any time is described by a ``typical state'' in the full Hilbert space, {so that the value of the von Neumann entropy for any subsystem is given by the Haar-averaged value}.  
The Page time is accordingly given by the time scale when $d_{B(t)} = d_{R(t)}$. 

While this derivation of the Page curve is largely kinematical and quasi-static, the {premise} that the states of the black hole plus radiation during the evaporation process can be considered ``typical'' is a highly non-trivial dynamical assumption. This assumption is plausible at a heuristic level, {and is partially supported by the conjecture that the evolution of the black hole subsystem should be governed by a highly chaotic and maximally scrambling Hamiltonian~\cite{Hayden:2007cs,susskind,butterfly}.}
But if a chaotic Hamiltonian evolution is indeed behind the emergence of the Page curve, one should be able to 
 directly identify the dynamical principles underlying it,  without invoking ``typicality of states.'' This is the main goal of the present paper.

We consider simple quantum dynamical toy models 
for an evaporating black hole and for an eternal black hole coupled to an infinite bath. 
In both cases, we assume that a chaotic Hamiltonian governs the black hole subsystem, and find that the Page curve results from simple universal features of operator evolution that are characteristic of a quantum chaotic system. 
 In particular, the change from increasing to decreasing behavior in the Page curve of the radiation may be understood from change of dominance between two distinct types of physical processes: ``continuous'' spreading of operators, and ``discontinuous'' void formation recently discussed in~\cite{Liu:2019svk}. This picture resonates well with and could provide a microscopic explanation for the recent semi-classical derivations 
of the Page curves in two-dimensional black hole systems~\cite{pen,Al1,Al2} (see also~\cite{Al3,
Rozali:2019day, 
Akers:2019nfi, 
Chen:2019uhq,
Al4, Penington:2019kki, 
Almheiri:2019qdq}), where~\eqref{avEn} arises from a switch in quantum extremal surfaces~\cite{netta}.  In particular, this suggests that  the ``island'' contribution in the derivation of the Page curve for the Hawking radiation in~\cite{Al2} may have a microscopic origin in void formation.

\begin{figure}[!h]
\begin{center}
\includegraphics[width=14cm]{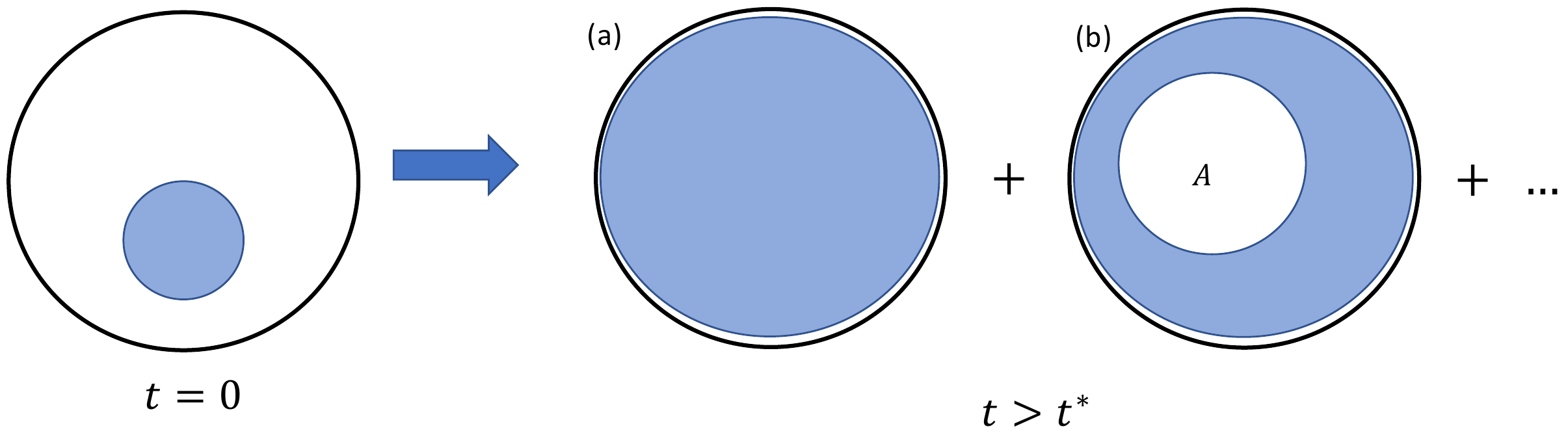}
\caption{Universal features of operator growth in a chaotic system. {The region within the black circle represents the space of degrees of freedom, and shaded regions indicate the subsystems where the operators are supported. A given initial operator evolves to a superposition of different final operators. After some time $t$ greater than the scrambling time $t^*$, a typical process is one where the operator becomes supported on the entire space, as shown in term (a). For any subsystem $A$, there is a small probability that the final operator is equal to the identity in $A$, as shown in the term (b). We refer to the presence of terms like (b) as void formation in $A$.}}
\label{fig:1d}
\end{center}
\end{figure}

The basic idea of our approach is as follows. Consider a quantum-mechanical system with an initial density operator $\rho_0 = \ket{\psi_0} \bra{\psi_0}$. To understand the  evolution of the entanglement of a subsystem $A$ with its complement $\bar A$, 
it is convenient to decompose $\rho_0$ into a basis of operators $\{\sO_\al \}$ which respect the tensor structure $\sH_A \otimes \sH_{\bar A}$, 
\be \label{opg}
\rho_0 = \sum_\al a_\al \sO_\al  \ .
\ee
The time evolution $\rho(t)$ of $\rho_0$ and its entanglement properties
can be obtained from the evolution of the set of operators $\sO_{\alpha}$ on the right-hand side of~\eqref{opg}\footnote{See~\cite{abanin,MS} for earlier discussion.}, which we refer to as an ``operator gas." 
For a chaotic system, the time evolution of a general operator 
exhibits certain universal behaviors, such as ballistic spreading and the decay of out-of-time-ordered correlation functions (OTOCs)~\cite{butterfly}, {which reflect the fact that any initial operator typically becomes supported in the entire system after a time scale known as the scrambling time, as shown in the (a) term in Fig.~\ref{fig:1d}.}
Another universal feature, recently identified in~\cite{Liu:2019svk} as being responsible for ensuring the unitarity constraint 
$S_A = S_{\bar A}$, is that an operator has a certain probability to develop a ``void," {as shown in the (b) term in Fig.~\ref{fig:1d}.} In this paper, we will show that void formation also underlies the Page curve and the transfer of information from the black hole to the radiation.

More explicitly, for any subsystem $A$ of the entire system, we can decompose the time evolution of an operator $\sO_\al$ in~\eqref{opg} as 
\be \label{yhe}
\sO_\al (t) =\sO_\al^{(1)} (t) + \sO_\al^{(2)} (t) , \qquad \sO_\al^{(1)} (t) = \tilde \sO_{\bar A} \otimes \bid_A
\ee
where $\bid_A$ denotes the identity operator in $A$, $\tilde \sO_{\bar A}$ is some operator  
in $\bar A$, and $\sO_\al^{(2)} (t)$ 
is an operator whose restriction to $A$ is orthogonal to $\bid_A$. {In Fig.~\ref{fig:1d}, the term (b) corresponds to $O_{\alpha}^{(1)}(t)$, and the term (a) is the largest contribution to $O_{\alpha}^{(2)}(t)$.}  Given that the space of all operators is a Hilbert space, we can also associate a weight or ``probability'' for $\sO_\al (t)$ to develop a void in subsystem $A$
\be 
P_{\sO_\al}^{(A)} (t) = {\Tr \le(\le(\sO_\al^{(1)} (t)  \ri)^\da\sO_\al^{(1)} (t)  \ri) \ov \Tr \le(\sO^\da_\al (t) \sO_\al (t) \ri) }  \ . \label{void_def}
\ee
Below we will refer to $P_{\sO_\al}^{(A)} (t) $ as the probability of forming a void in $A$.  The probability for a basis operator $\sO_\al$ to develop a ``macroscopic'' void is very small, exponentially suppressed by the 
number of degrees of freedom in the void region, but surprisingly can lead to $O(1)$ violation of unitarity when neglected~\cite{Liu:2019svk}.

\begin{figure}[!h]
\begin{center}
\includegraphics[width=15cm]{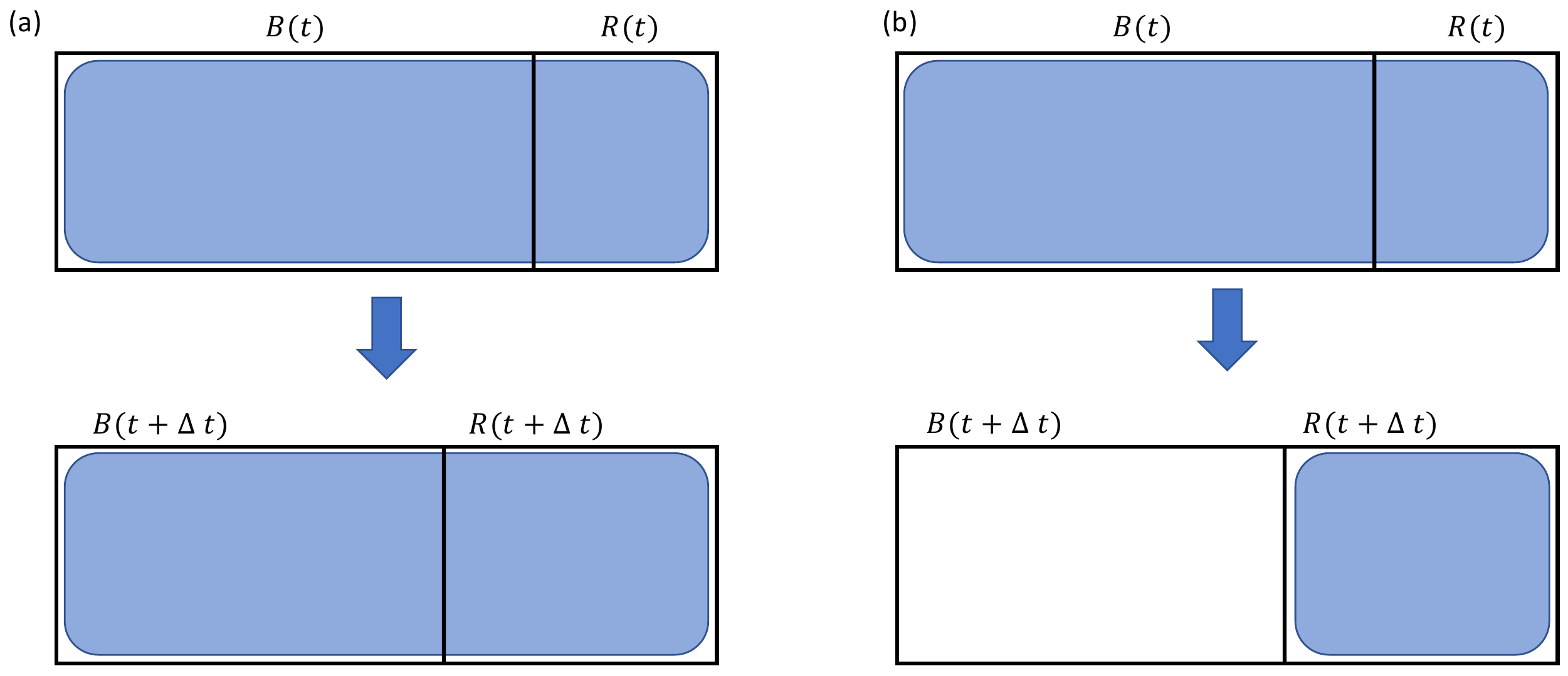} 
\caption{Evolution of the operator gas in an evaporating black hole. In all cases, the region on the left represents the black hole, and the region on the right represents the radiation. The black hole subsystem $B(t)$ grows smaller as a function of time during the evaporation process, while the radiation subsystem $R(t)$ grows larger. 
 (a) During evolution, an operator which is supported in the full system (i.e. including both $B$ and $R$) remains 
 supported in both subsystems. (b) Operators which are initially supported in the black hole at time $t$ have some probability of ``jumping'' outside $B(t+\Delta t)$ at a later time, forming a void which includes the black hole.}
\label{fig:bh}
\end{center}
\end{figure}

Applying the operator gas approach to a quantum model of black hole evaporation, one finds that 
the entanglement entropy\footnote{For simplicity, in this paper we will examine only the second Renyi entropy.} of the radiation can be separated into contributions from two distinct physical processes, illustrated in cartoon pictures in Fig.~\ref{fig:bh}, 
\be \label{enr} 
 e^{-S_2 ^{(R)}} 
 = \Tr \rho_R^2 =  e^{- \sS_{R (t)}} + e^{-\sS_{ B (t)}} + \cdots  \ 
\ee
where $\sS_B$ and $\sS_R$ were defined in~\eqref{avEn} and $\cdots$ denotes contributions which are suppressed by 
further powers of $e^{- \sS_R}$ or $e^{-\sS_B}$.\footnote{Except near $t=0$ or near the end of the evaporation, both $\sS_R$ and $\sS_B$ can be considered macroscopic.}
The first term on the right hand side of~\eqref{enr} 
arises from processes like the one shown in Fig.~\ref{fig:bh}(a), where a basis operator 
$\sO_\al$ originally in the black hole subsystem becomes supported in the black hole as well as the radiation subsystem due to the ``continuous'' process of Hawking radiation. Such processes would by themselves lead to indefinite growth of the entanglement entropy of the radiation, like in the  dashed line in Fig.~\ref{fig:page}. The second term in~\eqref{enr} comes from 
``discontinuous'' void formation processes illustrated in Fig.~\ref{fig:bh}(b), where the part of an operator originally supported in the black hole subsystem can ``jump'' to the radiation subsystem. Such processes are exponentially suppressed in terms of the coarse-grained black hole entropy before the Page time~\cite{page}, but dominate 
after the Page time and lead to the turn-around of the Page curve. 

The change of exponential dominance exhibited 
in~\eqref{enr} provides an explanation for why the Page curve is visible semi-classically, and is similar to the change of exponential dominance between two saddle points in the Euclidean path integral in \cite{Penington:2019kki,Almheiri:2019qdq}, 
where the second void formation term in~\eqref{enr} arises from replica wormholes. 
We emphasize that the change of exponential dominance in~\eqref{enr} in our analysis comes from dynamical processes and is Lorentzian in nature. 

\begin{figure} 
\includegraphics[width=16cm]{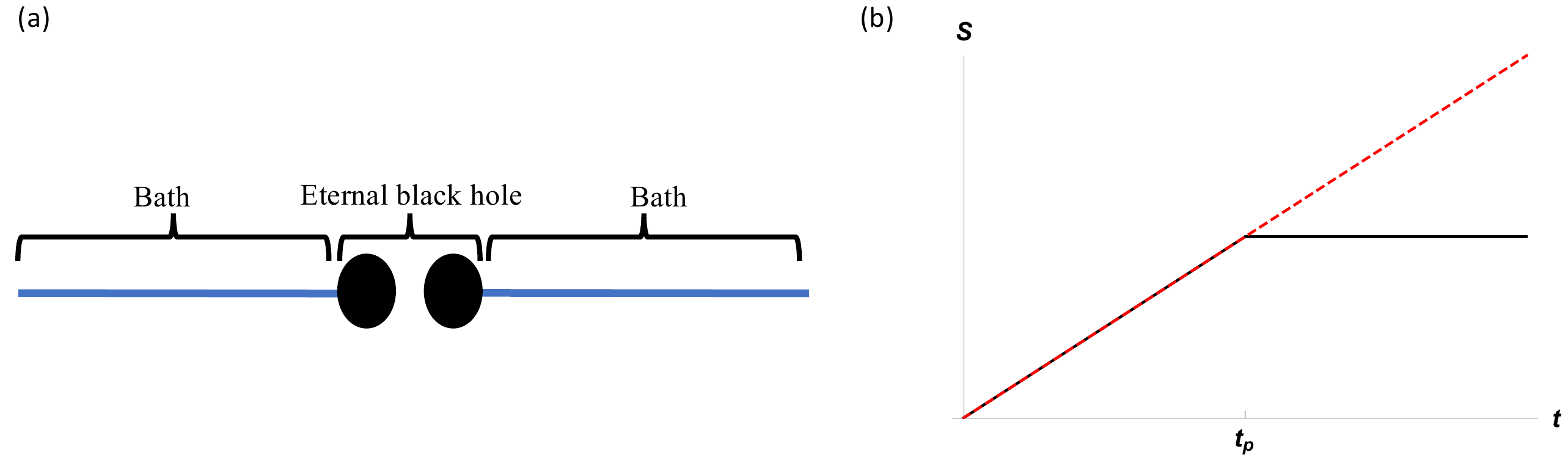} 
\caption{(a) A schematic illustration of a two-sided eternal black hole coupled to a one-dimensional bath. (b)~The evolution of the entanglement entropies of the black hole and the bath in this setup. The black curve represents the evolution of the entanglement entropy of both the black hole and the bath under unitary evolution, and is hence a counterpart of the Page curve for this setup. The red dashed line shows the evolution of the entanglement entropy of the radiation from naive semiclassical calculations, or without including void formation processes, which is a manifestation of the information loss problem in this setup.}
\label{fig:eBH}
\end{figure} 

We also consider a simple quantum-mechanical toy model for an eternal black hole coupled to a one-dimensional bath, motivated by the discussion of~\cite{Al3,Almheiri:2019qdq}. See Fig.~\ref{fig:eBH}(a).  In this case the Hilbert spaces of the black hole and the bath do not change, but the two subsystems  can exchange quantum 
information through their interactions. As a result, if we start with an unentangled state between the black hole and the bath, 
the entanglement entropy of both subsystems should increase with time, and eventually saturate at $2 \sS_{\rm BH}$,  the maximum possible entropy of the finite-dimensional black hole system.  $\sS_{\rm BH}$ is again the coarse-grained entropy for the black hole, which is now a constant with respect to time. See the solid curve in Fig.~\ref{fig:eBH}(b) for the evolution of entanglement entropy expected from unitarity in this setup.
The saturation time may be considered a counterpart of the Page time. Applying the operator gas approach to such a system, we find the entanglement entropy of the bath has the form 
\be \label{uene} 
e^{-S_2^{(\rm bath)}} =  e^{-a \seq t } + e^{- 2 \sS_{\rm BH}} + \cdots  
\ee
where $\seq$ is the equilibrium entropy density for the bath, and $a$ is some constant. 
The value of $a$ depends on the nature of the bath system and the initial state of the bath. For illustrations, we consider two models of bath systems,  a ``chaotic" bath, and a ``free'' bath. 
In~\eqref{uene}, the first term again arises from ``continuous spreading'' of operators as in Fig.~\ref{fig:bh}(a), while 
the second term comes from the ``discontinuous'' void formation of Fig.~\ref{fig:bh}(b), which becomes dominant after the 
counterpart of the Page time 
\be 
t_p = {2 \ov a} { \sS_{\rm BH} \ov  \seq}  \ .
\ee
Again the contribution of void formation ensures that the entanglement entropy of the bath is equal to that of the black hole after $t_p$, and hence plays the same role as replica wormholes and island contributions. Without this contribution, the entanglement entropy of the bath grows indefinitely, as shown in the dashed curve in figure \ref{fig:eBH}(b).

We also explore the dynamical mechanisms for the transfer of information 
between the black hole and the radiation/bath in both the evaporation and the bath models. In the  Hayden-Preskill protocol~\cite{Hayden:2007cs}, a process in which a message thrown into a black hole comes out in the radiation, one finds that again processes like Fig.~\ref{fig:bh}(b) are responsible for  transferring the information from the black hole to the radiation. If we ignore such processes, the information is simply lost from both subsystems.

The plan of the paper is as follows. In Sec.~\ref{sec:toy_model}, we consider a simple toy model for black hole evaporation, and derive the Page curve for the black hole and the radiation from simple assumptions about operator growth in a chaotic system. We then discuss the consequences of void formation for the Hayden-Preskill protocol in this model. In 
Sec.~\ref{sec:ebh}, we describe our models for an eternal black hole coupled to a bath, and explain the role of void formation in these models for both the emergence of the Page curve and information transfer. 
We conclude in Sec.~\ref{sec:conc} with some open questions and future directions. We have also included two technical Appendices to supplement the discussion of the main text.

\section{A toy model for black hole evaporation}
\label{sec:toy_model}

\subsection{The model and setup}  \label{sec:set}

In this section, we consider a simple quantum mechanical model for black hole evaporation, in which the degrees of freedom of a black hole system slowly go into a system of radiation. The time-evolution in the black hole is assumed to be {chaotic}.

We take the full quantum-mechanical system $L$ to consist of $k$ generalized ``spins'', each of which has a Hilbert space of dimension $q$.  The full Hilbert space is then $\sH = \otimes_{i=1}^k \sH_i$ with total dimension $q^k$, where $i$ labels different spins. $k$ is assumed to be very large. For computational convenience, we will take $q$ large in all subsequent sections, but 
our conclusions should {qualitatively} apply to any finite $q$.

The black hole and radiation subsystems at time $t$ are respectively denoted as $B(t)$ and $R(t)$, with 
$L = B(t) \cup R(t)$. 
Initially, $B(t=0)= L$, so the black hole consists of the full system, and $R(t=0)$ is empty.
The initial state  is taken to be a pure state.  

\begin{figure} 
\includegraphics[width=8cm]{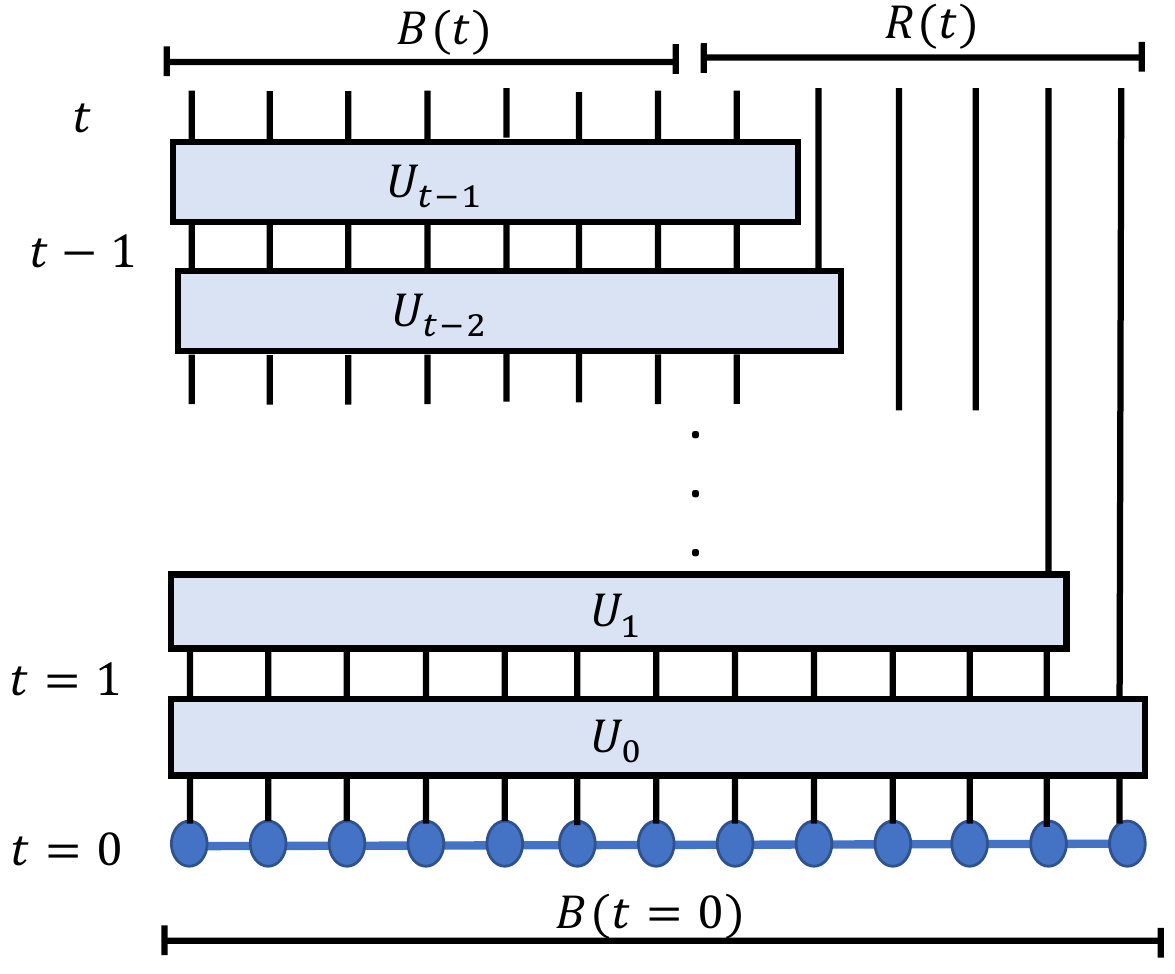}
\caption{Time-evolution of the evaporating black hole. In the first half of each time-step, a Haar-random matrix from $U(q^{k-t})$ is applied within $B(t)$, and in the second half, a site is taken out of $B$ and put in $R$ to give $B(t+1)$ and $R(t+1)$. }
\label{fig:evap_ev}
\end{figure}

We will take time steps to be discrete. 
The time-evolution from $t=n$ to $t=n+1$ consists of {first applying a unitary operator $U_t$ on}
the subsystem $B(t)$, and then taking one spin\footnote{Note that if instead we took some $n\ll k$ spins from the black hole into the radiation at each time-step, the dependence of the entanglement entropies on the coarse-grained entropies $\sS_{B(t)}$ and $\sS_{R(t)}$ at all times would be unaffected.} out of $B(t)$ and making it part of $R(t+1)$.
 At $t \geq k $, $B(t)$ is empty while $R(t) = L$. 
 The time-evolution is illustrated in Fig.~\ref{fig:evap_ev}. No non-trivial time-evolution is applied within $R (t)$.
We can interpret the length of each time step as being equal to the scrambling time of the black hole.\footnote{For a realistic black hole system, the scrambling time will change as the black hole evaporates, but this does not affect our conclusions.} We assume $U_t$ arises from a chaotic Hamiltonian, whose specific form will not be of concern to us. 
Our goal is to derive the Page curve for the second Renyi entropies $S_2$ for $B(t), R(t)$ using only some general properties of $U_t$.

To calculate  $S_2$ for 
 $B(t)$ and $R(t)$ using the operator gas approach, 
 it will be convenient to expand the  density operator $\rho(t)$ of the system 
 in terms of a complete set of basis operators which respects the tensor product structure $\sH_{B (t)} \otimes \sH_{R (t)}$ for 
 all $t$. More explicitly, for the $i$-th site (or spin), we define an orthonormal basis of operators $O_c^i$, $c= 0, ..., q^2-1$, which is normalized as 
\be \label{jen}
\tr ((O_c^i)^\da O_d^j)  = q \de_{cd}  \de_{ij}, \qquad O_0^i = \bid_i
\ee
where $\bid_i$ is the identity operator for the Hilbert space $\sH_i$. Orthogonality with $O_0^i$ implies $O_{c}^i, c =1, \cdots q^2-1$, are all traceless. A convenient choice of basis (suppressing indices $i$) 
is 
\be \label{obd}
O_c = X^{s_1} Z^{s_2}, \qquad s_1,s_2 = 0,1, \cdots q-1
\ee
where $X$ and $Z$ are the shift and clock matrices given more explicitly in Appendix~\ref{sec:op_app}. 
 An orthonormal basis of operators for the full system, which will be denoted as 
$\sO_\al, \al =0,1, \cdots q^{2k}-1$,  can be obtained from tensor products of $\{O_c^i\}$. 
The basis operators satisfy
\be \label{hne}
\Tr \sO_\al^\da \sO_\b = \de_{\al \b} ~ q^k  \ .
\ee
$\sO_0 = \bid$ is the identity operator for the full Hilbert space $\sH$, and all other $\sO_\al$'s are traceless.  

We will take the initial density operator to be a pure state. As explained in Appendix \ref{sec:op_app}, any pure state can be written in the form  
\be  \label{heno}
\rho_0  
= {1 \ov q^k}  \sum_{a \in I}  \mathcal{P}_a 
\ee
where $I$ is a set of $q^k$ mutually commuting operators $\{\sP_a\}$ (including the identity operator), which can again be normalized as in~\eqref{hne}.  
Under time evolution, we can expand $\sP_a(t)$ in the $\{\sO_{\alpha}\}$ basis,  
\be \label{evop}
\sP_a (t) = U^\da (t) \mathcal{P}_a U(t) =  \sum_\b c_a^\b (t) \sO_\b
\ee
where $U(t)$ denotes the evolution operator. Note that the identity operator remains the identity at all time. From unitarity of $U(t)$, 
\be 
\sum_\b |c_a^\b  (t) |^2  =1  \ .
\ee
We can interpret $|c_a^\b  (t) |^2$ as the probability of  $\mathcal{P}_a$ evolving to $\sO_\b$ at time $t$.

Under time-evolution, using~\eqref{evop},  the reduced density matrix for a subsystem $A$ is given by\footnote{For notational convenience we will take states of the system to evolve by $U^\da$, i.e. $\rho(t) = U^\da (t) \rho_0 U (t)$.} 
\be \label{dens1}
\rho_A (t) = {\rm Tr}_{\bar A}  \rho (t) = 
 {1 \ov q^k} \sum_{a \in I}{\rm Tr}_{\bar A} \mathcal{P}_a (t) =
{1 \ov q^{|A|}}  \bid_A + {1 \ov q^{|A|}} \sum_{a \in I} \sum_{\b \in A, \b \neq \bid_A}  c_{a}^\b (t) \sO_\b \ 
\ee
where $|A|$ is number of spins in $A$. 
Due to the tracelessness of all nontrivial basis operators, only $\sO_\al$ of the form $\sO_\b \otimes \bid_{\bar A}$ with $\sO_\b$ an operator in $A$ (denoted by $\b \in A$)  contribute to 
${\rm Tr}_{\bar A} \sO_\al$. 
Note that if none of the $\sO_\b$ on the right-hand side of~\eqref{evop} is contained entirely in $A$ at time $t$, then the only 
operator contributing to $\rho_A (t)$ is the identity operator, in which case the second term in~\eqref{dens1} is absent and $A$ is maximally entangled with $\bar A$.

Using \eqref{dens1}, we find that 
\be \label{eej}
e^{-S_2^{(A)} (t)} = {\rm Tr}_{A} \rho_{A}^2 (t) = \frac{1}{q^{|A|}} \sum_{a_1, a_2 \in I}\sum_{\beta\in A} c_{a_1}^{\beta} c_{a_2}^{\beta \ast}(t)  ~ \approx   {1 \ov q^{|A|}} + 
 {1 \ov q^{|A|}} N_A (t)
 \ee
 where\footnote{{We ignore the contribution from terms with $a_1 \neq a_2$ in the second-to-last expression in \eqref{eej}, as if we assume that the phases of $c^{\beta}_a$ are random in a chaotic system, then the total contribution from such terms is suppressed by order $O(q^{-|A|})$ relative to terms with $a_1 = a_2$.}} 
\be \label{ejn}
N_{A} (t) \equiv    \sum_{a \in I} \sum_{\b \in A, \b \neq \bid_A} | c_{a}^\b (t)|^2   
  \ee
is the expected number of nontrivial operators in the set $I$ which are ``localized'' in subsystem $A$ at time $t$. 

We will now assume that $U_t$ at each time step comes from a chaotic Hamiltonian. 
In~\cite{Liu:2019svk}, it was argued that a chaotic Hamiltonian can be characterized by its probability distribution for ``void formation," where the probability of forming a void in a subsystem $A$ is as defined in \eqref{void_def}. 
From studies of local random unitary circuits, it was conjectured there that in a chaotic system, after the scrambling time, the probability for a generic initial operator $\sO$ to develop a void in a sufficiently large subsystem $A$ is given by the``random void distribution":
\be \label{pgd}
P_{\sO}^{(A)}(t) = {1 \ov d_A^2} 
\ee
where $d_A$ is the dimension of the Hilbert space of $A$. In particular, for the initial operators $\sP_a$, \eqref{pgd} implies that 
\be
P_{\sP_a}^{(A)}(t)  = \sum_{\b \text{\, with void in $A$}} \le|c_a^\b (t) \ri|^2 = {1 \ov d_A^2} \ 
\ee
where $``\b \text{\, with void in $A$}"$ refers to the requirement that each $\sO_\b$ in the sum has the form $\sO_{\bar A} \otimes \bid_A$ with $\sO_{\bar A}$ some operator  
in the complement of $A$. \eqref{pgd} is the key property that we will use below to derive the Page curve.
We stress again that each time step in our model should be considered a unit scrambling time of the 
black hole subsystem. Hence, we will assume that \eqref{pgd} holds for each time-step and for any subsystem $A \subset B(t)$.

Note that one choice of time-evolution for which \eqref{pgd} can be readily seen to hold is when we take $U_t$ to be a Haar-random unitary from $U(q^{|B(t)|})$, where $|B(t)|$ is the number of spins in $B(t)$. But the random void distribution should apply to more general chaotic $U_t$.

\subsection{The Page curve for the radiation and void formation}   \label{sec:ra}
 
\begin{figure}[!h] 
 \includegraphics[width=7cm]{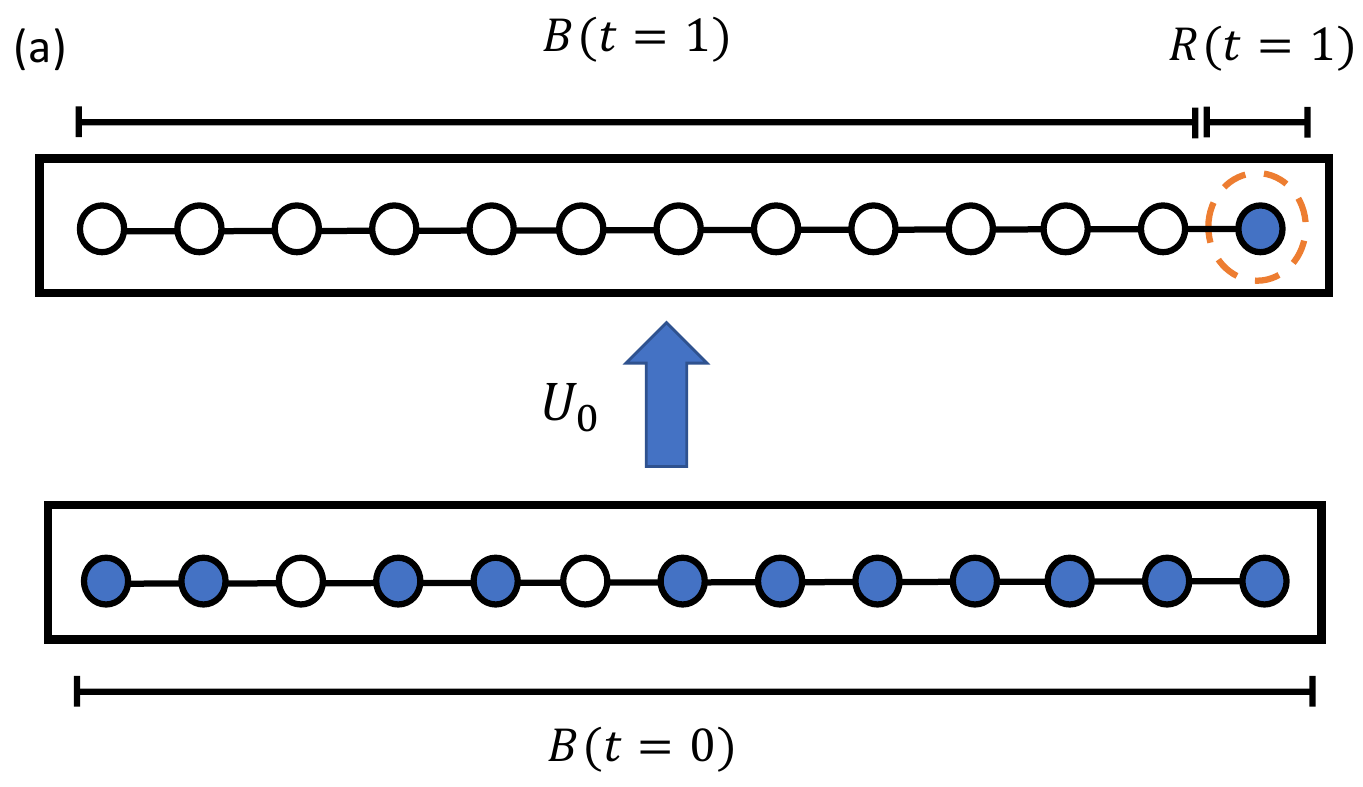} ~~~~~~ \includegraphics[width=7cm]{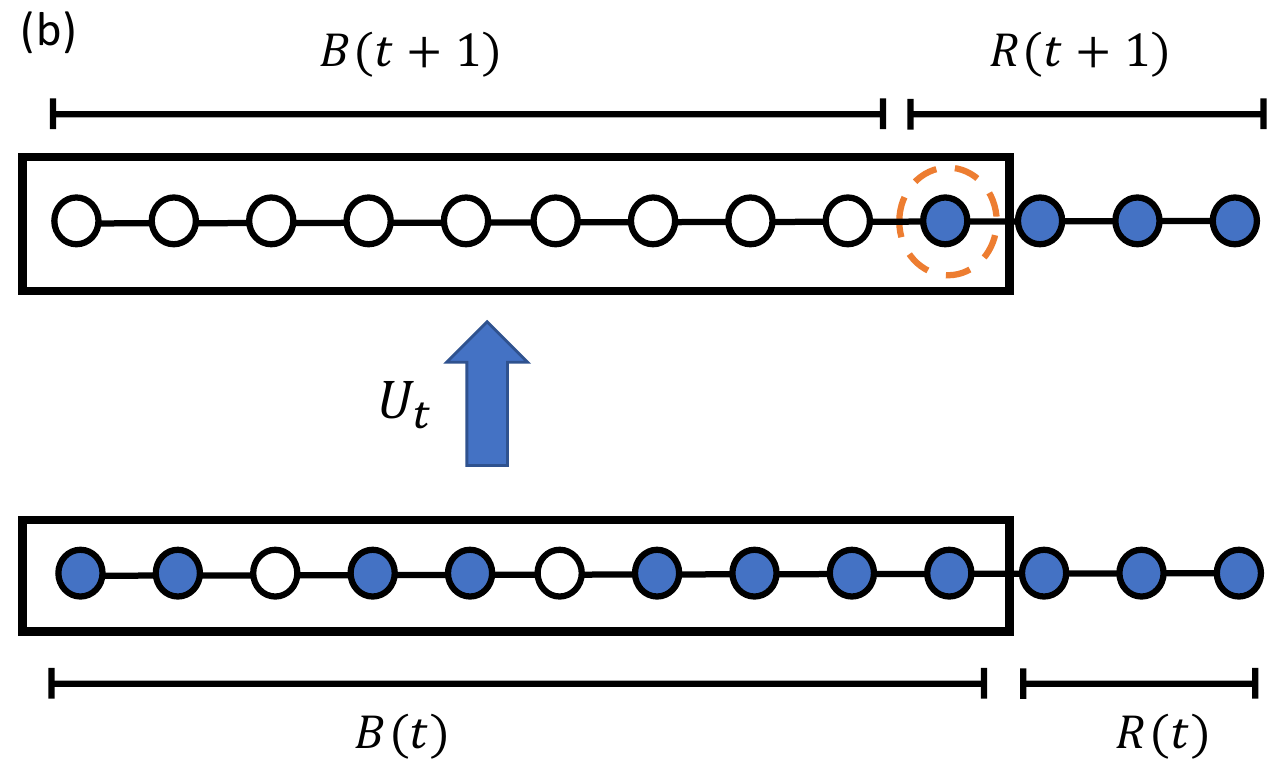}
 \caption{Macroscopic void formation processes in which operators with non-trivial support somewhere in $B(t)$ evolve to operators trivial at all sites in $B(t+1)$, shown in (a) at $t=0$ and in (b) at a later time.  Operators have non-trivial support at shaded sites, the region enclosed by the rectangle is the one where the time-evolution operator at time $t$ acts, and the encircled site is the Hawking radiation emitted between times $t$ and $t+1$.}
 \label{fig:voids}
 \end{figure}

Let us first consider the evolution of $S_2$ for the radiation subsystem, using~\eqref{eej} with $A = R(t)$.
For this purpose, we need to find the expected number $N_{R(t)}$ of nontrivial operators 
that are ``localized'' in $R(t)$. 
Consider first $t=1$. The only contribution to $N_{R(t)}$ comes from void formation processes in the black hole, where an initial operator $\sP_a$ 
transitions to an operator which is equal to the identity in $B(t=1)$, i.e. trivial at all sites except the one which will be taken as radiation, see Fig.~\ref{fig:voids}(a). For a single $\sP_a$, from the random void distribution~\eqref{pgd}, the probability of this process is $q^{- 2 (k-1)}$. Since the total number of initial operators $\sP_a$ is $q^k$, the total expected number $N_{R(t)}$ for all $\sP_a \in I$ is $q^k q^{- 2 (k-1)} = q^{-k+2}$. 
We then find that
 \be\label{eon0}
e^{- S_2^{(R (t=1))}} =  {1 \ov q} + q^{- k +1}  = e^{-\sS_{R(t=1)}} + e^{- \sS_{B(t=1)}} 
 \ee 
 where $\sS_{B (t=1)} = (k-1) \log q$ and $\sS_{R(t=1)} = \log q$ are the coarse-grained entropies for the black hole and the radiation. 
 Note that the dominant term comes from the identity in~\eqref{dens1}, and thus 
  $R(t=1)$ is close to maximally entangled with $B(t=1)$. 
 
 The story at subsequent time steps works similarly: the contributions to $N_{R(t)}$ come from 
processes of forming a void in $B(t)$. {The leading contribution in the large $q$ limit comes from processes where the void is formed} during the evolution from $t-1$ to $t$, 
\be 
\sO_{B(t-1)} \otimes \sO_{R(t-1)} \to \bid_{B(t)} \otimes  \sO_D \otimes \sO_{R(t-1)} = \bid_{B(t)} \otimes  \sO'_{R(t)} \label{big_void}
\ee
where $D$ denotes the Hawking radiation emitted from $t-1$ to $t$, and {$\sO_{B(t-1)}$ and $\sO_D$ are non-trivial.} {See Fig.~\ref{fig:voids}(b).}  
 From~\eqref{pgd}, such processes  
 give $N_{R(t)} =q^k q^{-2 (k-t)} =  q^{-k + 2t}$. Since $|R(t)| = t$,
 we thus find from~\eqref{eej} that 
 \be\label{yeh1}
e^{- S_2^{(R (t))}} =  {1 \ov q^t} + q^{- k +t} =  e^{-\sS_{R(t)}} + e^{- \sS_{B(t)}}    \ .
 \ee 
 The two terms change dominance at the Page time $t_p = {k \ov 2}$. Thus, before the Page time void formation processes are exponentially suppressed compared with the contribution of the identity operator in~\eqref{dens1}, but  they dominate after the Page time.

\begin{figure}[!h] 
 \includegraphics[width=7cm]{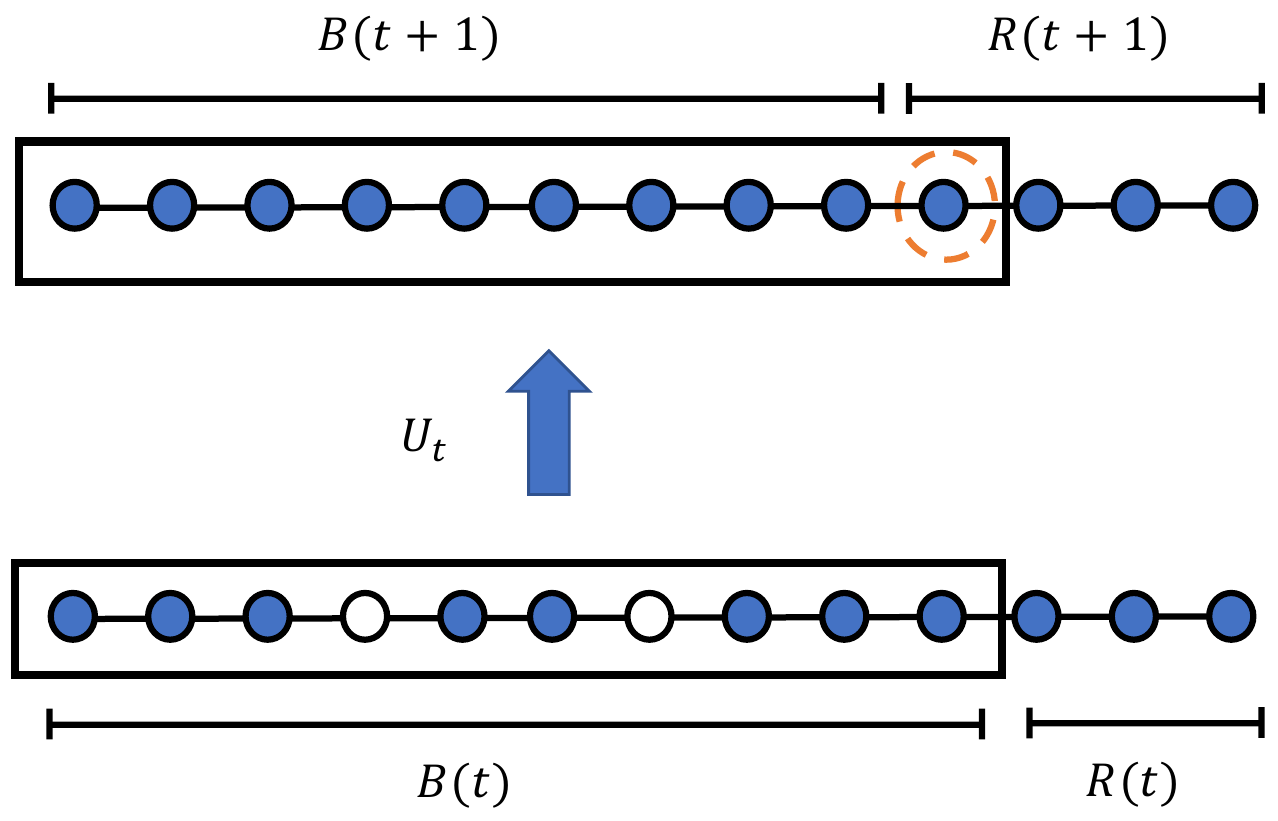} 
  \caption{A typical process in the chaotic evolution is one in which operators with non-trivial support somewhere in $B(t)$ evolve to operators with non-trivial support at all sites in $B(t)$.}
 \label{fig:typical_process}
 \end{figure}

Note that a typical process during the evolution of an operator is
 \be 
\sO_{B(t-1)} \otimes \sO_{R(t-1)} \to \sO_{B(t)} \otimes  \sO_D \otimes \sO_{R(t-1)} = \sO_{B(t)} \otimes  \sO'_{R(t)} \label{no_void} 
\ee
with $\sO_{B(t)}$ non-trivial at all sites in $B(t)$ and $\sO_D$ non-trivial, as shown in Fig.~\ref{fig:typical_process}. If all non-trivial operators evolved in this way at all times, then the only contribution to~\eqref{dens1} for $A=R(t)$  would be from the identity operator in the initial density matrix, and one would have only the first term in~\eqref{yeh1}, leading to indefinite growth of $S_2^{(R(t))}$ {continuing past the Page time}.  

In the derivation of the island formula for a toy model involving Jackiw-Teitelboim gravity with an end-of-the-world brane in \cite{Penington:2019kki}, $e^{-S_2^{(R)}}$ is seen to be a sum of exponentials coming from distinct saddle points in the Euclidean path integral, which have precisely the same form as~\eqref{yeh1}.   In our discussion, we are able to attribute the two contributions to distinct dynamical processes in operator growth. The discussion above also suggests that the contribution of the ``island'' is the semi-classical manifestation of void formation.  

We finally note that at finite $q$, the story should hold qualitatively except that the behavior of the system near the transition region at the Page time will be more complicated.

 \subsection{The Page curve for the black hole} \label{sec:bh} 

Let us now look at evolution of $S_2$ for the black hole, taking $A = B(t)$ in~\eqref{eej}. 
At $t=1$, only those $\sO_\b$ in~\eqref{evop} which have the identity operator at the site which is taken to be
$R(t=1)$ will contribute to $N_{B(t=1)}$. {In the large $q$ limit, the random void distribution~\eqref{pgd} can be applied to a single spin,} so the probability for a  single operator $\mathcal{P}_a$
to remain in $B(t=1)$ is $q^{-2}$.
 Thus the total expected number of operators that remain in $B(t=1)$ is given by 
 $N_{B} (t=1) = q^k q^{-2} = q^{k-2}$. Since $|B(t=1)| = k-1$, we thus find from~\eqref{eej} 
  \be \label{eon}
  e^{- S_2^{(B (t=1))} } = {1 \ov q^{k-1}} + {1 \ov q^{k-1}} q^{k-2}   =  {1 \ov q^{k-1}} + {1 \ov q}
    \ee 
{We can immediately see in the large $q$ limit that} equation~\eqref{eon} is identical to~\eqref{eon0} as required by 
  unitarity. But note that now the first term in~\eqref{eon}, which is equal to $e^{-\sS_{B(t=1)}}$, arises from the contribution of the identity operator instead of that of non-trivial operators. 
  
   \begin{figure}[!h] 
 \includegraphics[width=7cm]{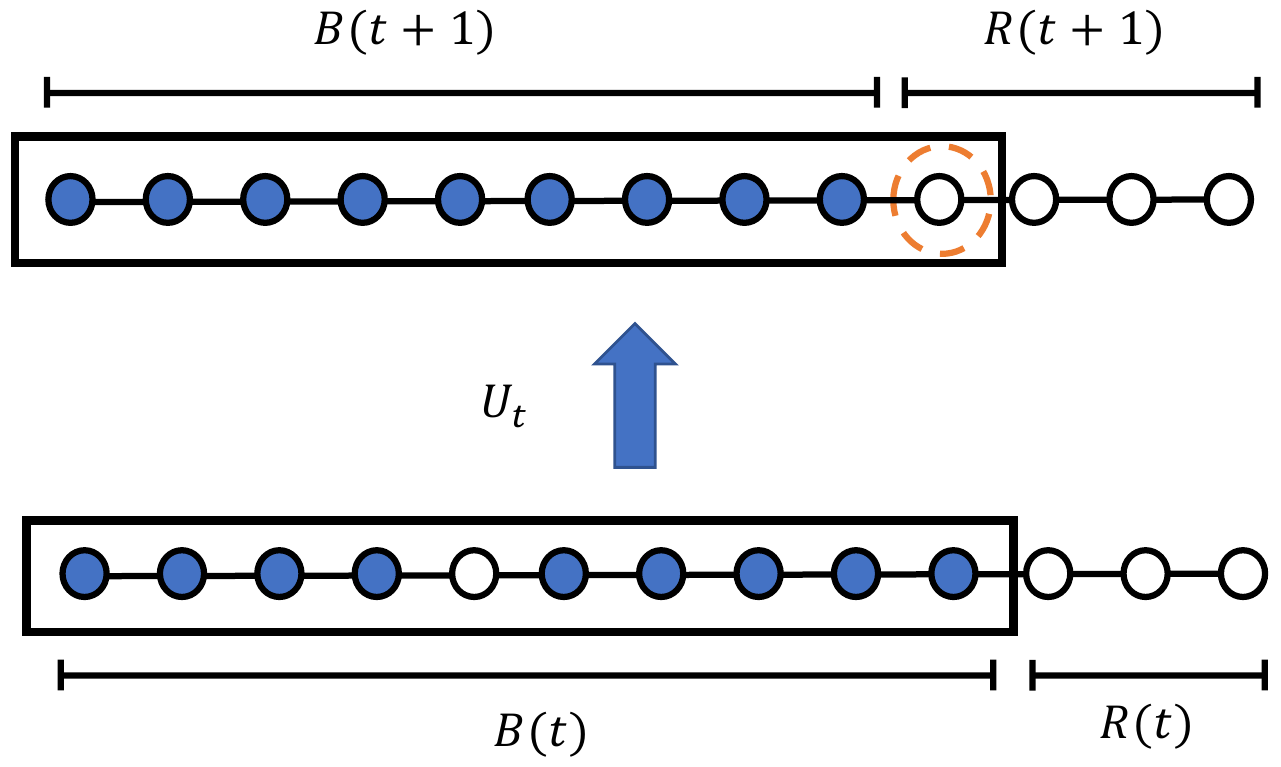} 
 \caption{Microscopic void formation processes in which operators with non-trivial support somewhere in $B(t)$ evolve to operators trivial at one site in $B(t+1)$, which becomes part of the Hawking radiation at time $t+1$. Such processes contribute to the second term in \eqref{yeh}.}
 \label{fig:small_voids}
 \end{figure} 
  
For general $t$ we have similarly  $N_B (t+1) = {1 \ov q^2} N_B (t)$ due to the probability 
$1/q^2$ of the process (shown in Fig.~\ref{fig:small_voids}) 
\be 
\sO_{B(t)} \otimes \mathbf{1}_{R(t)} \to \sO_{B(t+1)} \otimes  \mathbf{1}_D \otimes \mathbf{1}_{R(t)}  \label{small_void} 
\ee      
where $D$ is the Hawking radiation emitted from time $t$ to time $t+1$, {and $\sO_{B(t)}$ and $\sO_{B(t+1)}$ are non-trivial.}
  Since $|B(t)| = k-t$, we have 
  \be\label{yeh}
  e^{- S_2^{(B (t))} } = {1 \ov q^{k-t}} + {1 \ov q^{k-t}} q^{k-2t}   =  {1 \ov q^{k-t}} + {1 \ov q^t} 
  \ee 
  which again agrees with~\eqref{yeh1}. Now the first term, which comes from the identity operator, dominates after the Page time $t_p = {k \ov 2}$. As expected, this implies $B(t)$ becomes almost maximally entangled with $R(t)$ after the Page time.

  The change of dominance between the two terms in~\eqref{yeh} at the Page time $t_P = {k \ov 2}$ 
  may be seen as the microscopic origin of the change of dominance between two sets of quantum extremal surfaces 
  in the semi-classical discussion of~\cite{pen,Al1}. In that discussion, before the Page time, the quantum extremal surface 
  is trivial, reflecting the ``perturbative'' nature (that is, independent of the coarse-grained black hole entropy) of entanglement growth in~\eqref{yeh}, while after the Page time, the quantum extremal black hole is close to the black hole horizon 
  and given by the coarse-grained black hole entropy. The operator growth origin of~\eqref{yeh}  highlights that when the entanglement entropy  of the black hole is given by the coarse-grained entropy, it is close to being maximally entangled with the radiation, and  its reduced density matrix is close to the identity operator.

Note that in contrast to the discussion of the entanglement entropy for the radiation in the last subsection, which involves forming voids in the entire black hole subsystem (which is macroscopic at all times of interest), the above discussion of the evolution of the entanglement entropy for the  black hole only involves forming ``microscopic" voids at single sites with a probability independent of the coarse-grained black hole entropy. 
This may be viewed as a ``perturbative'' contribution in terms of the semi-classical gravity description.  
This may explain why a more conventional quantum extremal surface prescription was sufficient for correctly calculating the Page curve for the black hole~\cite{pen,Al1}, while a new element involving ``islands'' had to be introduced to calculate the Page curve for the radiation~\cite{Al2}.

\subsection{Information transfer from the black hole to the radiation} 

Let us now use the operator gas approach to study the Hayden-Preskill process~\cite{Hayden:2007cs}
and explore how information originally in a black hole is transferred to the radiation through the Hawking process. 
In our setup, this can be achieved by taking a subspace $P$ consisting of $p \ll k$
spins in the black hole to be maximally entangled with a $q^p$-dimensional reference system $Q$, and studying the evolution of the mutual information of various subsystems with $Q$ at later times. 
We can introduce the reference system either for a young black hole, i.e. at $t=0$, or for an old black hole after the Page time. 
We will see in both cases that by maintaining unitarity, void formation ensures that the information originally in subsystem $P$ is actually transferred to the radiation. Neglecting void formation, one finds that the information is simply lost. 

\subsubsection{A young black hole} 

At $t=0$, we take $p$ out of $k$ spins to be maximally entangled with a $q^p$-dimensional reference system $Q$. 
The time evolution operator is $U_L\otimes \mathbf{1}_Q$, where $U_L$ is the time-evolution operator for 
$L = B \cup R$ as described in Sec.~\ref{sec:set}.  We then examine the time-evolution of the mutual information of $Q$ with $B(t)$ and $R(t)$ to track the information that was originally contained within $P$.

We take the initial state to have the form 
\be 
\ket{\psi_0} = \ket{\chi}_{PQ} \otimes \ket{\phi}_{L-P}  
\ee
where $\ket{\chi}_{PQ}$ is a maximally entangled state between $P$ and $Q$, and $\ket{\phi}$ is an arbitrary pure state. 
As explained in Appendix \ref{sec:op_app}, the initial density operator can then be expanded in terms of basis operators as 
\be 
\rho_0 = {1 \ov q^{k+p}} \sum_{i} \sO_i^Q \otimes \tilde \sO_i^P \otimes \sum_{\al \in I}  \sP_a^{L-P} 
\ee
where $i$ goes over {\it all} basis operators in system $Q$, $\tilde \sO_i^P$ is fixed from $\sO_i^Q$,  and $I$ is a set 
of $q^{k-p}$ commuting operators $\{\sP_a\}$ of $L-P$.  In particular, when $\sO_i^Q$ is given by $\bid_Q$ (say for $i=0$),
the corresponding $\tilde \sO_0^P$ is given by $\bid_P$ and vice versa.\footnote{See Appendix~\ref{sec:op_app} for more details on the operator form of maximally entangled states.} The density operator at time $t$ is then given by 
\be \label{trg}
\rho (t) = {1 \ov q^{k+p}} \sum_{i} \sO_i^Q \otimes U^\da \le(\tilde \sO_i^P \otimes \sum_{a \in I}  \sP_a^{L-P} \ri) U \ .
\ee

Since $U$ does not act on $Q$, $Q$ is maximally entangled with $L = B \cup R$ at any $t$, i.e. 
\be 
\rho_Q = {1 \ov q^p} \bid_Q \quad \implies \quad S_2^{(Q)} = p \log q \equiv \sS_Q  \ .
\ee
From unitarity, the mutual information of $Q$ with $B(t)$ and $R(t)$ should satisfy\footnote{$I_2 (A,B)$ below is the second Renyi version of the mutual information: $I_2 (A, B) = S_2 (A) + S_2 (B) - S_2 (AB)$.}
\be \label{eun1}
I_2 (Q, B(t)) + I_2 (Q, R(t)) = 2 S_2^{(Q)}= 2 \sS_Q 
\ee
at all times. At $t=0$, $I_2 (Q, R) =0$ and all the information of the subsystem $P$ is in $B$. 

The calculations of various quantities $S_2^{(B)} (t), S_2^{(QB)} (t)$ and $S_2^{(R)} (t), S_2^{(QR)} (t)$, that are 
needed to obtain the mutual information between $Q$ and $B, R$ are  in parallel with our earlier 
discussion of Sec.~\ref{sec:bh} and Sec.~\ref{sec:ra}, so we will only briefly mention the calculation of $S_2^{(R)} (t)$ and $S_2^{(QR)} (t)$ as illustrations.   

To obtain $\rho_R$ we need to take the trace over $Q$ and $B(t)$. {When tracing over $Q$,  only the $i=0$ term 
in~\eqref{trg} corresponding to the identity operator contributes, i.e. }
\be 
\rho_R (t) =  {1 \ov q^{k}} {\rm Tr}_{B(t)} \le( U^\da \le(\bid_P \otimes \sum_{\al \in I}  \sP_a^{L-P} \ri) U \ri) \ .
\ee
{When tracing over $B(t)$, as in the discussion of Sec.~\ref{sec:ra}, only operators with a void in subsystem $B(t)$
can contribute.} The only difference here from the discussion of Sec.~\ref{sec:ra} is that we now start with a more restricted set of $q^{k-p}$ 
operators, which gives 
 \be\label{yehn}
e^{- S_2^{(R)} (t) } =  {1 \ov q^t} + q^{- k - p +t  }  + \cdots  
 \ee  
where the first term comes from the identity and the second comes from void formation in $B(t)$. 
For $S_2^{QR} (t)$, since now $Q$ is part of the subsystem, any $i$ in~\eqref{trg} contributes, so we have 
$q^{k+p}$  initial operators, which then gives 
 \be\label{yehn2}
e^{- S_2^{(QR)} (t)} =  {1 \ov q^{t + p}} + q^{- k  +t  }  + \cdots   \ .
 \ee  

We thus find that the Renyi mutual information between $Q$ and $B$ evolves as 
 \be 
 I_2 (Q, R; t) = \begin{cases} 
0 & t<(k-p)/2 \\
 (p-k + 2t) \log q & (k-p)/2< t < (k+p)/2 \\
 2 \sS_Q  & t >  (k+p)/2
 \end{cases}
 \label{i1}
 \ee
where we have only kept the leading term in the large $q$ limit. 

From an analysis similar to that of Sec.~\ref{sec:bh}, we find  
\be 
 I_2 (Q, B; t) = \begin{cases} 
2 \sS_Q  & t<(k-p)/2 \\
 (k+p - 2t) \log q & (k-p)/2< t < (k+p)/2 \\
0  & t >  (k+p)/2
 \end{cases} \ .
 \label{i2}
 \ee
We see that~\eqref{i1} and~\eqref{i2} indeed satisfy~\eqref{eun1}. 

From~\eqref{i2} we see that the information starts ``leaking'' out of the black hole at $t = {k-p \ov 2}$ when 
$|Q| + |R (t)| = |B(t)|$, and the information will have completely left at $t = {k + p \ov 2}$ when $|B(t)| + |Q| = |R (t)|$.
Between these two time scales, the information is shared between the black hole and radiation.  

Without including void formation processes in $B(t)$, we would have $S_2^{(Q \cup R)}(t) = (t+p) \log q$, $S_2^{(R)}(t) = t \log q$ 
and $I_2 (Q, R; t) =0$ for all $t < k$, while $I_2 (Q, B; t)$ is still given by~\eqref{i2}. We would then find that the information 
leaves the black hole, but does not show up in the radiation, and is thus lost.

\subsubsection{Old black hole and secret sharing}  

We now briefly discuss the story of an old black hole as in the original Hayden-Preskill protocol~\cite{Hayden:2007cs}, 
{but instead of taking the evolution of the black hole by a random unitary, we only assume that it is a chaotic evolution obeying the random void distribution~\eqref{pgd}. }

Consider an old black hole $B$ which is maximally entangled with a radiation system $R$ with $|R| > |B|$.
One adds to the black hole a system $A$ representing a diary thrown into it and the combined 
system $\tilde B = A \cup B$ is acted on by a unitary $U$. After the action of $U$, we separate from $\tilde B$ a subsystem $D$, which is the newly emitted radiation. We will denote the remaining black hole subsystem as $B'$, so that 
$\tilde B = D \cup B'$, and the full radiation as $R' = D \cup R$. 
A main point of~\cite{Hayden:2007cs} was that the
information of $A$ can be obtained from $R' = D \cup R$ with significant probability {if $d_D \gg d_A$, where $d_{D,A}$ are respectively the dimensions of the Hilbert space of $D$ and $A$.}  We again maximally entangle $A$ with a reference system $Q$ and track the flow of information from system $A$ using the mutual information of $Q$ with various subsystems. 


We will see below that void formation is again responsible for ensuring the information originally in system $A$ is indeed  
transferred to the full radiation subsystem $R'$. In fact we will see that the secret in $A$ is not in any of $B', D, R$ subsystems
alone, but can be recovered by having any two of them. The technical details are again very similar to those of previous sections and we will be brief.

The state of the full system after we throw in the diary has the form 
\be 
\rho_i = \rho_{QA} \otimes \rho_{BR} 
\ee
where $\rho_{QA}$ is the density operator for a maximally entangled state between $Q$ and $A$, and $\rho_{BR}$ 
is the density operator for a maximally entangled state between $B$ and $R$. They can be written respectively as ({see Appendix \ref{sec:op_app}})
\bega 
\rho_{QA} = {1 \ov d_A^2}  \sum_i \tilde \sO^Q_{i} \otimes \sO^A_i , \qquad 
\rho_{BR} = {1 \ov d_B^2}  \sum_\al \sO^B_{\al} \otimes \tilde  \sO^R_\al   \ .
\end{gather} 
Note that the  $\tilde \sO^R_\al$ are not basis operators for the entire system $R$, but instead for some $d_B$-dimensional subspace of $R$ that is maximally entangled with $B$. The final state has the form 
\be \label{ftp}
\rho_f = {1 \ov d_A^2 d_B^2} \sum_\al \sum_i  \tilde \sO^Q_{\al} \otimes  U (\sO^A_\al \otimes  \sO^B_{i} ) U^\da \otimes \tilde  \sO^R_i  
\ee
where $U (\sO^A_\al \otimes  \sO^B_{i} ) U^\da$ can then be further separated into a sum of products of basis operators of $
B'$ and $D$.

Now let us consider the reduced density matrices for various subsystems in the final state. 
Since $U$ does not act on $Q$ and $R$, these subsystems are still maximally entangled with their respective complements, i.e.   
\be 
\rho_Q = {1 \ov d_A} \bid_Q , \qquad \rho_R = {1 \ov d_B} \Pi   
\ee   
where $\Pi$ is the projector onto the subspace of $R$ which is maximally entangled with $B$. 
Since $Q$ is maximally entangled with the combined system $L = A \cup B \cup R = B' \cup D \cup R$, its mutual information with any subsystem $C$ of $L$ and its complement $\bar C$ in $L$ satisfies 
\be \label{eun}
I (Q, C) + I(Q, \bar C) = 2 \log d_A  \ . 
\ee

Now consider $\rho_{B'}$, which receives contributions from operators in~\eqref{ftp} of the form $\sO_a^{B'} \otimes \bid_{Q} \otimes \bid_D \otimes \bid_R$.
This is only possible when both $\al$ and $i$ are zero in~\eqref{ftp}. Similarly for $\rho_D$, $\rho_{DB'}$, and $\rho_{QR}$.  We thus find that these density matrices are all maximally mixed 
\be 
\rho_{B'} = {1 \ov d_{B'}} \bid_{B'}, \quad  \rho_{D} = {1 \ov d_{D}} \bid_{D}, \quad 
\rho_{DB'} = {1 \ov d_A d_B} \bid_D \otimes \bid_{B'} , \quad  \rho_{QR} = {1 \ov d_A d_B} \bid_{Q} \otimes \Pi \ .
\ee

Now consider $\rho_{DR}$ which can be written as 
\be 
\rho_{DR} = {1 \ov d_A d_B^2} \sum_\al {\rm Tr}_{B'} \le(U \bid_A \otimes \sO_\al^B U^\da \ri) \otimes \tilde \sO_\al^R
= {1 \ov d_B d_D} \bid_{DR} + \tilde \rho_{DR}
\ee
where the nontrivial contribution $\tilde \rho_{DR}$ (the part not including the identity) comes from void formation in $B'$, i.e. the part of  $U \le(\bid_A \otimes \sO_\al^B \ri)  U^\da$ containing operators of the form $ \bid_{B'} \otimes \sO^D$. 
Similarly, the nontrivial part of $\rho_{QD}$ arises from the part of  $U \le(\sO^A_i \otimes \bid_B \ri)  U^\da$ containing $ \bid_{B'} \otimes \sO^D$. One can similarly find the non-trivial parts of the other reduced density matrices. 

The final results for $d_D \gg d_A$ are:
\be 
I_2 (Q, B') = {d_A^2 \ov d_D^2} \approx 0 , \quad I_2 (Q, R) = 0, \quad I_2 (Q, D) = {d_D^2 \ov d_B^2} \approx 0
\ee 
and 
\be 
I_2 (Q, DR) = 2 \log d_A - {d_A^2 \ov d_D^2}  , \quad I_2 (Q, B'D) = 2 \log d_A, \quad I_2 (Q, B'R) = 2 \log d_A - {d_D^2 \ov d_B^2} \ .
\ee 
So the relation between $Q$ and various subsystems have the structure of secret sharing among three parties $B', D, R$. In particular, $I_2 (Q, DR) \approx 2 \log d_A$ corresponds to the fact that the information can be recovered from the radiation when $d_A/ d_D\ll 1$. {Without void formation in $B'$,} $I_2 (Q, DR)$ would be zero at all times.

\section{An eternal black hole coupled to an infinite bath} \label{sec:ebh}

We will now consider a toy model for an eternal black hole coupled to 
a bath in $(1+1)$-dimensions recently discussed in~\cite{Al3} (see also~\cite{Al4,Almheiri:2019qdq,Mathur:2014dia}).  Again we will find that void formation is responsible for the emergence of the Page curve shown in Fig.~\ref{fig:eBH}, and the transfer of information from the black hole to the bath.

\subsection{Description of the model and setup} 

\begin{figure}[!h]
\begin{center}
\includegraphics[width=14cm]{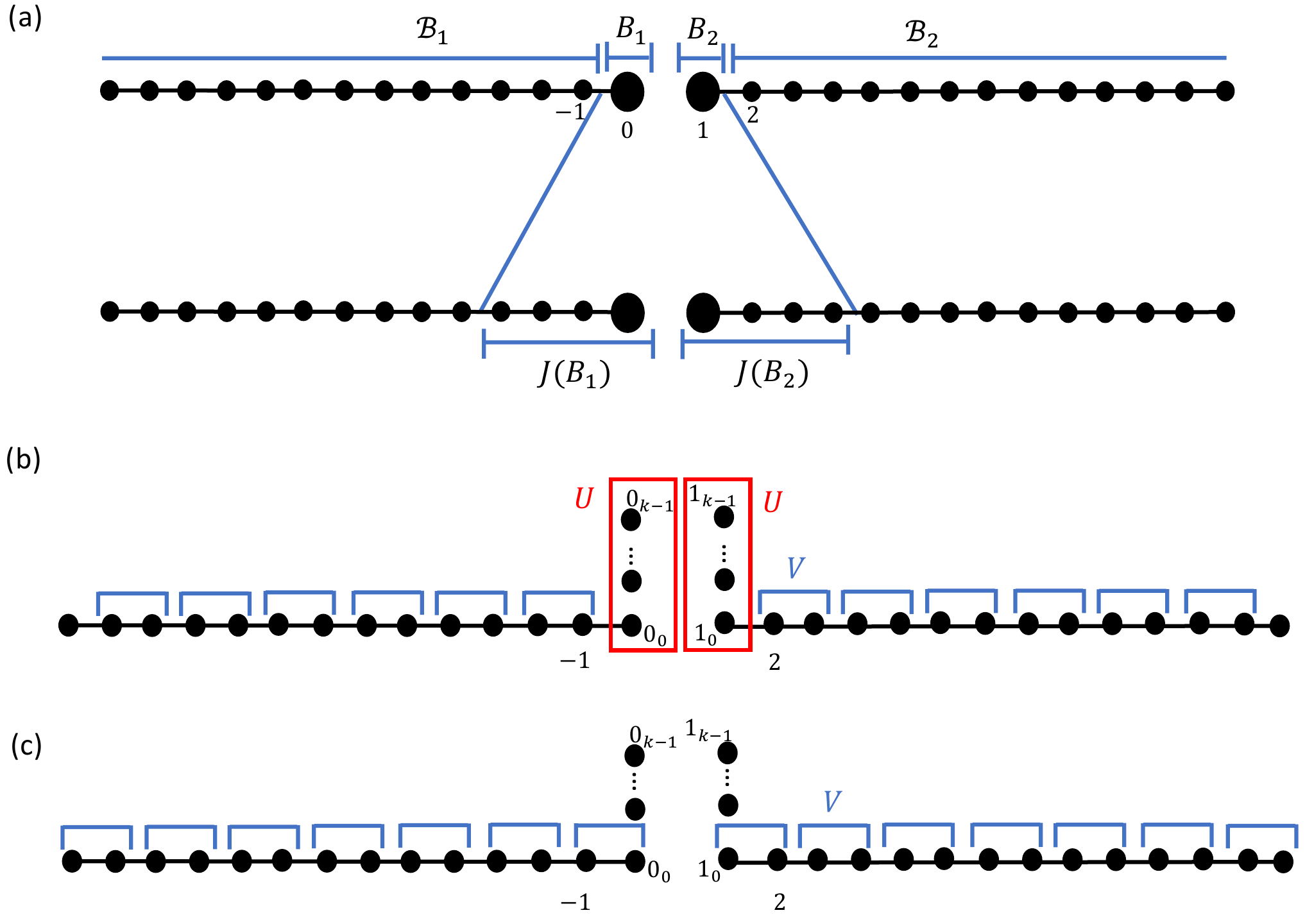} 
\caption{Lattice toy model for an eternal black hole coupled to an infinite bath. The subsystem at two sites, $0$ and $1$, describes the black hole. Various regions of interest are shown in (a). 
(b) and (c) show the time evolution. 
We imagine that site $1$ consists of $k = \log_q N  \gg 1$ $q$-spins which interact among one another, and one of them, $1_0$ is coupled to $2$. Similarly with site $0$. 
Unitary evolution at even time-steps is given by (b), where we apply {some unitary matrix $U$} {that is assumed to come from a chaotic Hamiltonian} within sites 0 and 1, and unitary matrices $V$ between 2 and 3, -1 and -2, and so on.  At odd time-steps, shown in (c), we apply unitaries $V$ to sites $0_0$ and $-1$, $-2$ and $3$, $1_0$ and $2$, $3$ and $4$, and so on. Note that $V$ between different sites at different times can be different; we use the same symbol for notational convenience.} 
\label{fig:bhm}
\end{center}
\end{figure}

Consider a (1+1)-dimensional lattice system extending from $-\infty$ to $\infty$, where the local Hilbert spaces of sites other than 0 and 1 have dimension $q$, while the local 
Hilbert spaces at sites $0$ and $1$ have dimension $N= q^k$, 
with $k$ large. The quantum subsystems at $0$ and $1$ are taken in a thermal field double state at infinite temperature, and describe an eternal black hole.\footnote{The subsystem at $0$ and $1$ can be viewed as the $(0+1)$-dimensional boundary dual for a $(1+1)$-dimensional black hole. It can also be 
viewed as a toy microscopic description of the black hole with details about the spacetime structure suppressed.}
There is no interaction between sites $0$ and $1$, but there are ``internal'' interactions 
at sites $0$ and $1$ respectively. 
The spin chain at the remaining sites corresponds to an infinite $(1+1)$-dimensional non-gravitational system, the ``bath,'' which the black hole couples to. We will take the bath to have $L$ sites on each side and take $L$ to infinity at the end.  
The dimension of the Hilbert space for the full system is thus $q^{2k + 2L}$. 
 A $q$-dimensional subspace at site $0$ is coupled to site $-1$, and a $q$-dimensional subspace at site $1$ is coupled to site $2$, to introduce interactions between the black hole and the bath. 
 The interactions among all the other sites are assumed to be local. 
{See Fig.~\ref{fig:bhm} for the configuration and details on time evolution.}  Note that the time-evolution does not couple the parts $[-\infty, 0]$ and $[1, \infty]$ of the system, which we will sometimes refer to as the left and right or $L$ and $R$ subsystems.

We take the initial state at $t=0$ to be
\be 
\rho_0 = \otimes_{i \leq -1} \rho_i \otimes \rho_{01} \otimes_{i \geq 2} \rho_i 
\label{BH_initial}
\ee
where $\rho_{01}$ is the density operator for the maximally entangled pure state between $0$ and $1$,
\be \label{urn}
\rho_{01} = \ket{\psi_{01}} \bra{\psi_{01}} , \qquad \ket{\psi_{01}} = {1 \ov \sqrt{N}} \sum_{n=1}^N \ket{n}_0 \ket{n}_1 
\ee
and $\rho_i = \ket{\psi} \bra{\psi}$ is a pure state which we will take to be the same for all sites of the bath. 
We can then expand $\rho_0$ in terms of basis operators discussed around~\eqref{jen}--\eqref{hne} as 
\be 
\rho_0 = {1 \ov q^{2L+ 2k}}  \sum_c O_c \otimes \tilde O_c \otimes \sum_{b \in I_{\rm bath}} \sO_b
\equiv  {1 \ov q^{2L+ 2k}}  \sum_{a \in I} \sO_a
\label{initial_rho}
\ee
where $O_c$ runs over all basis operators at site $0$ (with $\tilde O_c$ at site $1$ fixed by $O_c$), $I_{\rm bath}$ denotes the set of $q^{2L}$ basis operators formed by taking tensor products of all possible powers of the $Z_i$ operators defined in Sec.~\ref{sec:set} at different sites, and $I$ collectively denotes 
the whole set of initial operators. See Appendix~\ref{sec:op_app} for more details on how to obtain~\eqref{initial_rho}.  Note that the $\sO_a$ satisfy an orthonormality condition similar to~\eqref{hne}, with $q^k$ in~\eqref{hne} replaced by
$q^{2L + 2k}$.  

We are  interested in the evolution of $S_2$ for the black hole $B=B_1 \cup B_2$ and bath $\sB = \sB_1 \cup \sB_2$ subsystems. See Fig.~\ref{fig:bhm}(a). Instead of~\eqref{BH_initial}, we can also consider an initial state in which the left and right bath systems are entangled with each other. This will not lead to any difference in the behavior of entanglement growth for the regions $B$ and $\sB$.\footnote{It will, however, change the evolution of the mutual information between $\sB_1$ and $\sB_2$, as we discuss in section \ref{sec:mutual}.} 
{Recall from~\eqref{eej}--\eqref{ejn} that $S_2$ for a subsystem $A$ is given by}
\be \label{hjn}
e^{-S_2^{(A)} (t)}  =  {1 \ov d_A } + 
 {1 \ov d_A}N_A (t), \quad   N_{A} (t) \equiv    \sum_{a \in I} \sum_{\b \in A, \b \neq \bid_A} | c_{a}^\b (t)|^2
 \ee
 where $d_A$ is the dimension of Hilbert space in subsystem $A$.

 The qualitative features of our discussion will not be sensitive to the details of the unitary operator 
$U$ which governs the evolution of the black hole subsystem or the interactions $V$ among bath degrees of freedom
or  between the black hole and the bath, see Fig.~\ref{fig:bhm}(b)-(c).  We will assume $U$ is governed by 
some chaotic Hamiltonian such that under its action, a generic operator obeys the random void distribution~\eqref{pgd}
for any subsystem of $0$ or $1$. 
A solvable explicit example is to take $U$ to be a Haar-random unitary from $U(q^k)$ with $q^k$ large.

We will consider two types of $V$. We first consider a case where $V$ arises from a chaotic local Hamiltonian, and assume that 
under time evoltuion, a generic operator has the following properties: 
  \ben 
 \item In the systems $L \cup 0_0$ and $R\cup 1_0$, we have the property of sharp-light cone growth: an operator with endpoints $a$ and  $b$, with $a<b$, evolves into operators with end points $a-t$ and $b+t$ with total probability $1$. When one of the endpoints of an operator reaches the edge of either system (i.e. the black hole), it continues to grow only on the other end. 
We show in Appendix \ref{sec:light_cone} that for large $q$,  the sharp light-cone growth can be derived by applying the random void distribution to the action of each $V$.
 
 \item {The probability for an operator $\sO$ to develop a void in a subsystem $A$ lying within its light-cone 
 obeys~\eqref{pgd}.}
  For an operator that has reached the edge of either $L$ or $R$, the entire black hole site can be seen as a region ``within the lightcone" for the above statement. This can be seen as a consequence of the fact that the dynamics in the black hole are also chaotic. 
 \een
We will refer to such a bath as a ``chaotic bath.'' An explicit example is to take each $V$ in 
Fig.~\ref{fig:bhm}(b)--(c) to be an independent Haar-random matrix from $U(q^2)$ with $q$ large. 

{Another case we consider is one where $V$ models an integrable system. An example is to take all} $V$'s to be the same and have the form 
\begin{equation} 
V_{i' j', i j } = \delta_{i'j} \delta_{j'i}
\label{qp_u}
\end{equation} 
where $i$ and $j$, $i'$ and $j'$ label the basis of states at the two adjacent sites which are coupled by $V$. 
We will refer to the bath described by such $V$'s as a ``free'' bath.

 Below, as an illustration, we will mainly use the example of a chaotic bath. The analysis is very similar to that of Sec.~\ref{sec:toy_model} and Sec. II of \cite{Liu:2019svk}. Also, for simplicity, we will take $q$ to be large.  The results for a free bath will be discussed in Sec.~\ref{sec:free}.

\begin{figure} 
\includegraphics[width=14cm]{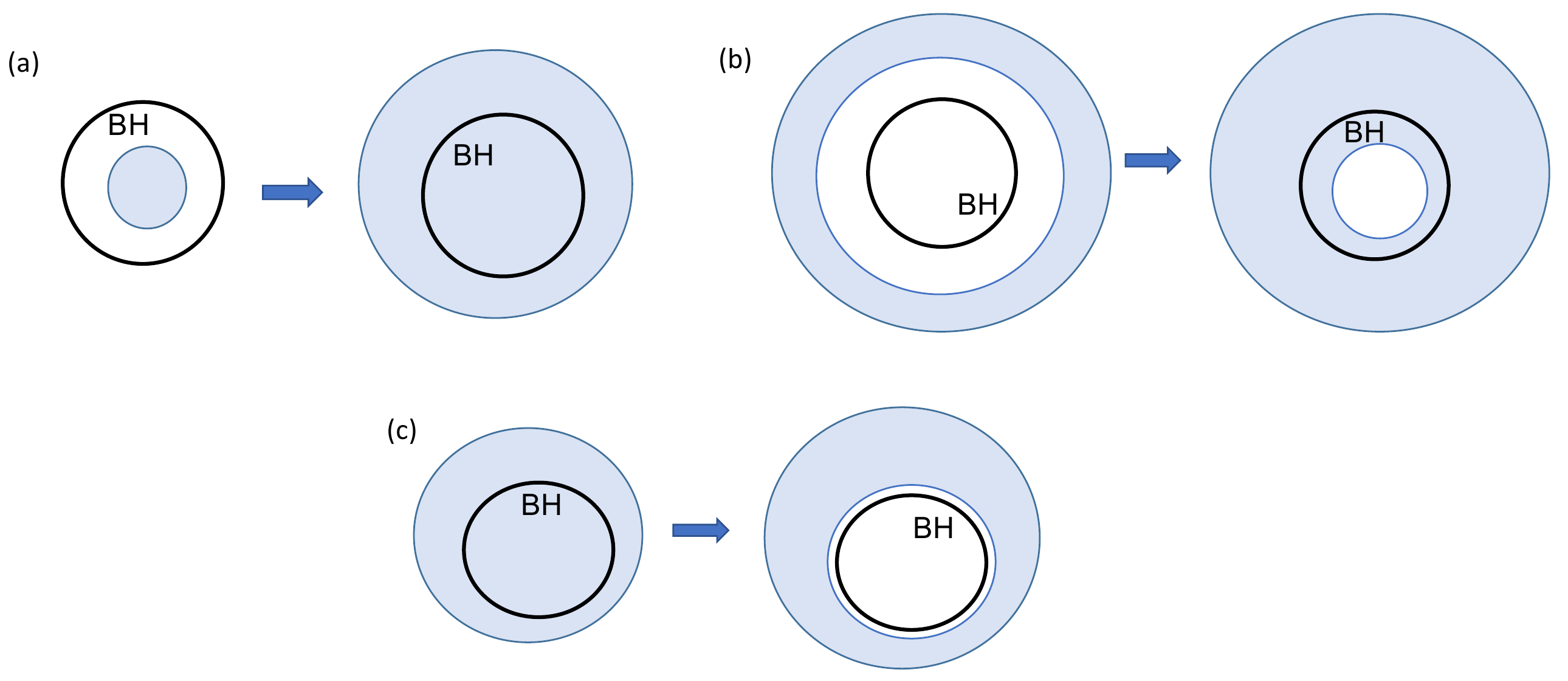}
\caption{Evolution of a black hole and the bath surrounding it, with the region inside the black circle representing the degrees of freedom of the black hole, and the region outside representing those of the bath. The shaded region denotes the support of an operator. 
 (a) During evolution, the operators making up the density operator of a black hole can ``leak'' outside the black hole. (b) Bath operators can ``fall'' into the black hole. 
(c)  Operators with support in the black hole region can also ``jump'' outside the black hole, forming a void which includes the black hole. 
}
\label{fig:bh_bath}
\end{figure}

We will see that $S_2^{(B)}$ grows due to processes like the one shown in Fig.~\ref{fig:bh_bath}(a), where operators grow out of the black hole, and finally saturates when only the identity operator remains in the black hole. $S_2^{(\sB)}$ grows due to processes like the one shown in Fig.~\ref{fig:bh_bath} (b). We will see that without the contribution from void formation processes shown in Fig.~\ref{fig:bh_bath}(c), the entanglement entropy of region $\sB$ appears to grow indefinitely, leading to $S_{\sB} \neq S_B$ after the entropy of the black hole saturates. This is precisely the version of the black hole information paradox revealed by the naive gravity calculations of \cite{Al3} when the island contribution is not correctly taken into account. 



\subsection{Evolution of entanglement of the black hole and the bath}

Let us first consider the black hole subsystem $B$. To find~\eqref{hjn}, we need to find the expected number $N_B (t)$ of 
operators from the initial set $I$ which remain in $B$ at time $t$. From sharp light-cone growth in the bath, an operator which originally has support outside subsystem $B$ will continue 
to have support outside $B$, and thus will never contribute to $N_B (t)$. At $t=0$, we have $N_B (t=0) = q^{2 k} -1 = d_B-1 \approx d_B$ as all the operators in $\rho_{01}$ are inside $B$. Now due to sharp light cone growth, among the operators inside $B$ at time $t$, operators with support at $0_0$ and $1_0$ will grow out of $B$ at any step where $B$ interacts with the bath. As a result, if $t$ is an even time just after an interaction between the black hole and the bath has taken place, then we can relate $N_B(t)$ to $N_B(t-2)$ in the following way: 
\be
N_{B}(t) 
=   N_B(t-2)q^{-4}
\label{nb_decrease}
\ee
where $q^{-4}$ is the probability of an operator being trivial at sites $0_0$ and $1_0$ after the chaotic unitary is applied within the black hole, obtained by applying the random void distribution within the black hole.

We thus find (we consider $t$ large so as not to be concerned with lattice effects)
that $N_B (t) = q^{2k  - 2t} $ 
and
\be \label{uenl}
e^{- S_2^{(B)}} = q^{-2 \seq t} + e^{- 2 \sS_{\rm BH}} \quad \implies \quad S_2^{(B)} =  \bca 2 \seq  t  & t < k \cr 
                        2 \sS_{\rm BH}  & t > k
                        \eca \ 
\ee
where we have introduced 
\be 
\sS_{\rm BH} = \log N = k \log q , \qquad s_{\rm eq} = \log q \ .
\ee
$\sS_{\rm BH}$ is the coarse-grained entropy for the black hole and $s_{\rm eq}$ is the ``equilibrium'' entropy density of the bath. 
Note that for $t >k $, all the nontrivial operators originally localized in subsystem $B$ have 
expanded outside $B$, and $S_2$ is given by the first term in~\eqref{hjn} (coming from the identity operator in $B$).  
 The processes underlying~\eqref{uenl} are illustrated in the cartoon picture of Fig.~\ref{fig:bh_bath}(a). 

To find $S_2^{(\sB)}$, we again consider equation~\eqref{hjn}, now with $A =\sB$. 
In this case, due to the light cone structure of the time-evolution, the expected number of operators in $\sB$ factorizes as~\cite{Liu:2019svk} 
\be \label{uen} 
N_{\sB} (t) =  q^{|\sB|- 2 t}  N (\sB, J(B); t), 
\ee
where the first factor comes from those operators which are inside $\sB$ at $t=0$ and remain in $\sB$ at time $t$. 
From the sharp light cone growth in the bath, these operators must lie outside the region $J(B) \equiv J(B_1)\cup J(B_2)$ indicated in Fig.~\ref{fig:bhm}(a) at $t=0$. 
The second factor $N(\sB, J(B), t)$ is the expected number of operators in region $J(B)$--indicated in Fig.~\ref{fig:bhm}(a)--
that transition to subsystem $\sB$. Such transitions can take place when the initial operators develop a void in $B$.

\begin{figure}[!h]
\includegraphics[width=8cm]{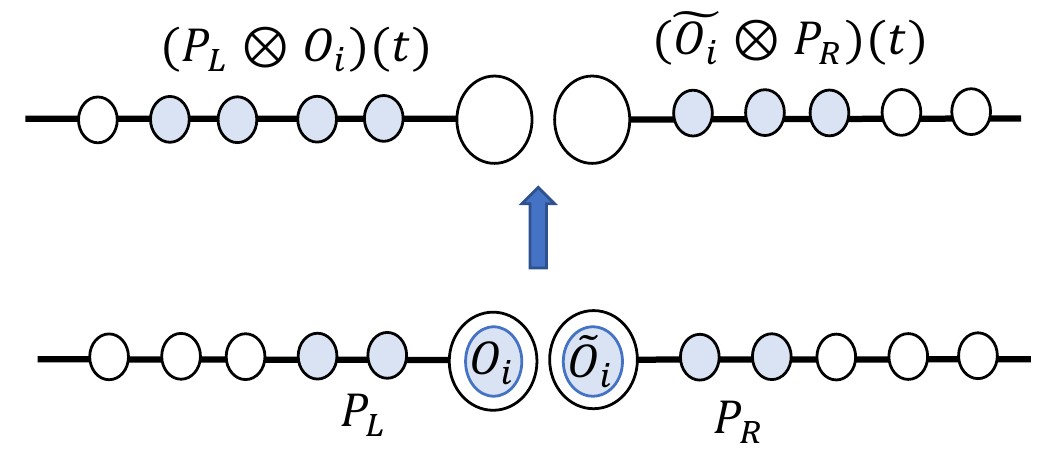}
\caption{{Processes of forming a void in $B$, which result in final operators contained in $\sB$.}  }
\label{fig:voia}
\end{figure}

The factor $q^{|\sB|- 2t}$ gives a contribution $2 t \, \seq$ to $S_2^{(\sB)}$, corresponding to processes where operators from increasingly distant regions from the black hole can expand into the 
black hole. Such processes, shown in figure \ref{fig:bh_bath}(b), will increase the entropy of $\sB$ indefinitely. 
Unitarity is restored when such processes are compensated for
by the process of void formation depicted in Fig.~\ref{fig:bh_bath}(c). This is captured by the factor of  $N(\sB, J(B), t)$.  
From the random void distribution~\eqref{pgd} we have 
\be  \label{hdf}
N(\sB, J( B), t) = 1 + {1 \ov d_B^2} N^2 q^{2t}  = 1 +q^{2t- 2k} 
\ee
where $d_B = N^2 $, and  $N^2 q^{2t}$ is the number of initial basis operators in $\rho_0$ 
in the region $J(B)$~\footnote{More explicitly, the factor of $\frac{1}{d_B^2}$ comes from applying the random void distribution for the final $U$ and $V$ acting before time $t$.}.  In~\eqref{hdf}, the first term comes from the identity operator in $J(B)$ and 
the second term from void formation of nontrivial operators.
The void formation processes contributing to~\eqref{hdf} are shown schematically in Fig.~\ref{fig:voia}.

Combining~\eqref{uen}--\eqref{hdf} and using~\eqref{hjn} for $\sB$, we 
find that $S_2^{(\sB)}$ precisely coincides with the expression for $S_2^{(B)}$ in~\eqref{uenl}. In particular, the entanglement entropy of the bath saturates for $t \geq t_p$ 
where  
\be \label{pagt}
t_p =k = \log_q N = {\log N \ov \log q} = {\sS_{\rm BH} \ov \seq} 
\ee
can be considered as the counterpart of the Page time.  

To conclude this subsection, let us make some remarks in connection with the gravity discussion of~\cite{Al3}: 

\ben

\item The time scale~\eqref{pagt} coincides with the semi-classical gravity estimate. In that context, $\seq = {c \ov \b}$, the entropy density for a CFT at inverse temperature $\b$. 


\item The transition from the first to the second line of~\eqref{hdf} at $t = t_p$ can be interpreted as a change of dominance between two sets of processes:  operator growth without void formation and with void formation. Before the Page time, void formation is exponentially suppressed, but its contribution becomes exponentially large after the Page time. 
This matches well with the gravity description, where unitarity is maintained by a jump in quantum extremal surfaces from surfaces without  ``island'' contributions to surfaces with ``islands.'' 

\een

\subsection{Evolution of mutual information between different parts of the bath} 
\label{sec:mutual}

It was pointed out  in~\cite{Liu:2019svk} that an immediate implication of void formation {in the systems considered there} is the generation of mutual information 
between regions which are separated by a void. We now examine the evolution of the mutual information between the two sides of the bath system, $\sB_1$ and $\sB_2$. We will first examine the story for the state~\eqref{BH_initial} where $\sB_1$ and $\sB_2$ are not entangled initially, and then comment on the case where $\sB_1$ and $\sB_2$ are maximally entangled in the 
initial state.

\begin{figure}[!h]
\includegraphics[width=8cm]{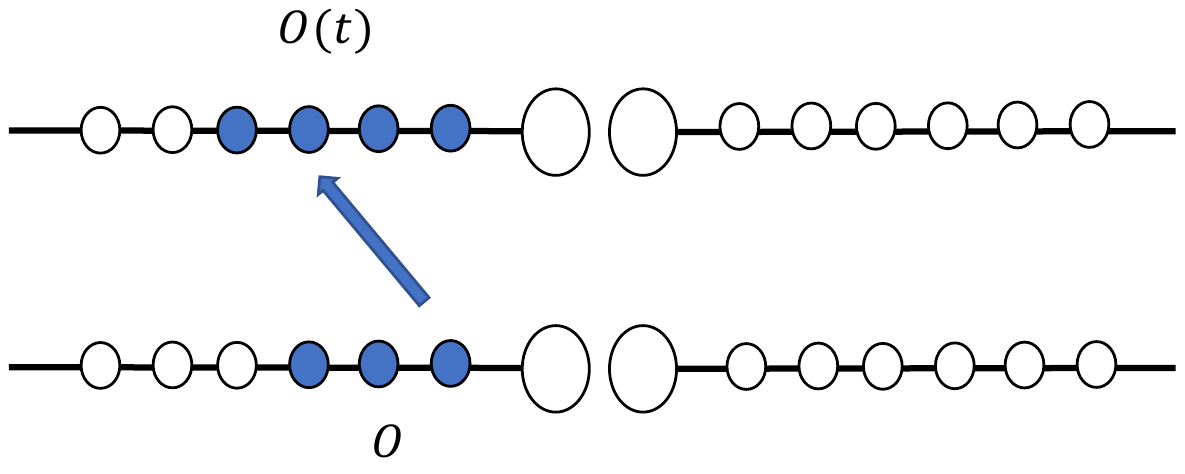}
\caption{{Processes of forming a void in $B_1$ which result in final operators contained in $\sB_1$.} }
\label{fig:voib}
\end{figure}

To find the evolution of mutual information between $\sB_1$ and $\sB_2$ in $\rho_0$ defined in~\eqref{BH_initial}, we only need to find $S_2^{(\sB_1)}$, as $S_2^{(\sB_2)}$ is identical due to the reflection symmetry, and $S_2^{(\sB)}$ was worked out in the last subsection. $S_2^{(\sB_1)}$ can be immediately found in close analogy with~\eqref{uen},
\be \label{gbe}
S_2^{(\sB_1)} = {1 \ov d_{\sB_1}}  \le(1 + N_{\sB_1} \ri) , \quad N_{\sB_1} = q^{|\sB_1|-  t}  N (\sB_1, J(B_1); t), 
\ee
where the first factor in $N_{\sB_1}$ again comes from the number of initial basis operators in $\sB_1$ which remain in $\sB_1$, and 
$N (\sB_1, J(B_1); t)$ gives the number of operators in $J(B_1)$ which can transition to $\sB_1$, i.e. by developing a void in $B_1$. {Note that in order for a final operator to be contained entirely in $\sB_1$, i.e. trivial in $B_1, B_2$ and $\sB_2$, 
it can only result from an initial operator which is trivial in the black hole subsystem.\footnote{Due to maximal entanglement between subsystems $0$ and $1$, if $\sO_i$ in~\eqref{initial_rho} is nontrivial, the corresponding $\tilde \sO_i$ must also be nontrivial. Since only the identity operator can evolve to the identity operator, we conclude that any operator which is nontrivially supported in the black hole subsystem cannot evolve into an operator which is nontrivial only on one side.}
 Thus the number of initial basis operators contained in $J(B_1)$ which can contribute to $N (\sB_1, J(B_1); t)$
is $q^{t}$.}  From the random void distribution~\eqref{pgd},  we have (in complete analogy to~\eqref{hdf}) 
\be \label{na1}
N(\sB_1, J(B_1); t) = 1 +  q^{t}  d_{B_1}^{-2}  = 1 + q^{t-2k}
\ee
{See Fig.~\ref{fig:voib} for processes contributing to~\eqref{na1}.}

We thus find 
\be 
S_2^{(\sB_1)}(t) =  \begin{cases} 
\seq t  & t<2k\\
2 \sS_{\rm BH}    & t\geq 2k  
\end{cases}
\ee
and as a result 
\be 
I_2(\sB_1, \sB_2; t) = \begin{cases} 0 & t<k  \\ 
2(t-k) \seq & k \leq t < 2k \\
2 \sS_{\rm BH}  & t\geq 2k 
 \end{cases}  \ . \label{mutual_B}
\ee 
We thus find that the mutual information between $\sB_1$ and $\sB_2$ starts growing at the Page time $t_p$ and 
saturates at $2 t_p$. 

The behavior~\eqref{mutual_B} can also be directly understood in a simple way from void formation in various regions. 
From~\eqref{hjn},~\eqref{uen}, and~\eqref{gbe}, we have\footnote{Note that the first term in~\eqref{hjn} for 
$\sB_{1,2}, \sB$ can be neglected as all these subsystems have an infinite dimensional Hilbert space.}
\bea 
I_2(\sB_1, \sB_2; t) &= & \log\le(\frac{N_{\sB}(t)}{N_{\sB_1}(t) N_{\sB_2}(t)}\ri) 
\cr
&= &   \log\le(\frac{q^{|\sB_1| + |\sB_2| -2t}N({\sB}, J(B); t)}{q^{|\sB_1| - t}N(\sB_1, J(B_1); t)~~q^{|\sB_2| - t} N(\sB_2, J(B_2); t)}\ri)  \cr
&=&   \log\le(\frac{N(\sB, J(B); t)}{N(\sB_1, J(B_1); t)N(\sB_2, J(B_2); t)}\ri)  \ .
\label{mutual_n}
\eea
Thus  the mutual information $I_2$ between $\sB_1$ and $\sB_2$ is a measure of void formation processes resulting in operators in $\sB$ which cannot be seen as a combination of independent void formation processes which would result in operators contained only within $\sB_1$ or only within $\sB_2$. The processes contributing to the upstairs and downstairs of~\eqref{mutual_n} were shown respectively in Fig.~\ref{fig:voia} and Fig.~\ref{fig:voib}.

Before the Page time $t<k$,  $N(\sB_1, J(B_1); t)$, $N(\sB_2, J(B_2); t)$ and $N(\sB, J(B); t)$ are all approximately $1$, and the mutual information is $0$. During the period $k<t<2k$, the processes of forming a void in $B$ give a significant contribution to $N(\sB, J(B); t)$, while 
processes of forming a void in $B_1$ or $B_2$ are still suppressed in $N(\sB_1, J(B_1); t)$ and $N(\sB_2, J(B_2); t)$. The mutual information between the regions increases linearly during this time.  For $t>2k$, all three quantities $N(\sB_1, J(B_1); t)$, $N(\sB_2, J(B_2); t)$ and $N(\sB, J(B); t)$ become exponentially large, and the time-dependence cancels between upstairs and downstairs of \eqref{mutual_n}. However, there is still a constant ratio by which the two quantities differ, corresponding to the fact initial operators non-trivial in the black hole subsystem cannot contribute to the 
denominator  (recall the discussion before~\eqref{na1}). The mutual information saturates at the log of this ratio.

Now let us consider the case where the initial state is maximally entangled between $\sB_1$ and $\sB_2$, i.e. 
\be \label{tau_initial}
\tau_0 = \rho_{01}\otimes_{i\geq2} \tau_{i, -i+1}
\ee 
where $\tau_{i, j}$ is a maximally entangled state between sites $i$ and $j$, 
\be 
 \tau_{i, j} =  \ket{\psi_{ij}}\bra{\psi_{ij}}, \quad \ket{\psi_{ij}} = \frac{1}{\sqrt{q}} \sum_{n=0}^{q-1} \ket{n}_i\ket{n}_j . 
\label{tau_ij}
 \ee  
This initial state can be expanded in terms of operators as 
\be \label{ej}
\tau_0 =  \frac{1}{q^{2k+2L}} \sum_{\alpha} \sO_{\alpha}^L \otimes \widetilde{\sO}_{\alpha}^{R} 
\ee
where $\alpha$ runs over all basis operators in the left system, and $\widetilde{\sO}_{\alpha}^{R}$ is determined by $\sO^L_{\alpha}$. In this case, the expressions for $S_2^{(B)}$ and $S_2^{(\sB)}$ are the same as~\eqref{uenl}.  
But $\sB_1$ is maximally entangled with the rest of the system at all times, as throughout the evolution there 
is no nontrivial operator that can result from~\eqref{ej} which is localized only in $\sB_1$. Thus    
\be
S_2^{(\sB_1)} = |\sB_1| \seq \ . 
\ee
and we find that
\be 
I_2(\sB_1,\sB_2; t) = \begin{cases} |\sB_1| + |\sB_2|- 2t \seq & t< k \\ |\sB_1|+ |\sB_2|- 2 \sS_{\rm BH} & t\geq k \end{cases} 
\ee
i.e. the mutual information between $\sB_1$ and $\sB_2$ initially decreases, and then saturates to a constant value after the Page time at which void-formation processes become dominant in $S^{(\sB)}_2$. 
 The above expression has a simple interpretation. $\sB_1$ and $\sB_2$ are maximally entangled initially. Increasing the entanglement of both subsystems with the black hole decreases the mutual information between them, until the time when the entanglement with the black hole saturates.

\subsection{Transfer of information between the black hole and the bath}\label{sec:trans}

\begin{figure}[!h]
\includegraphics[width=12cm]{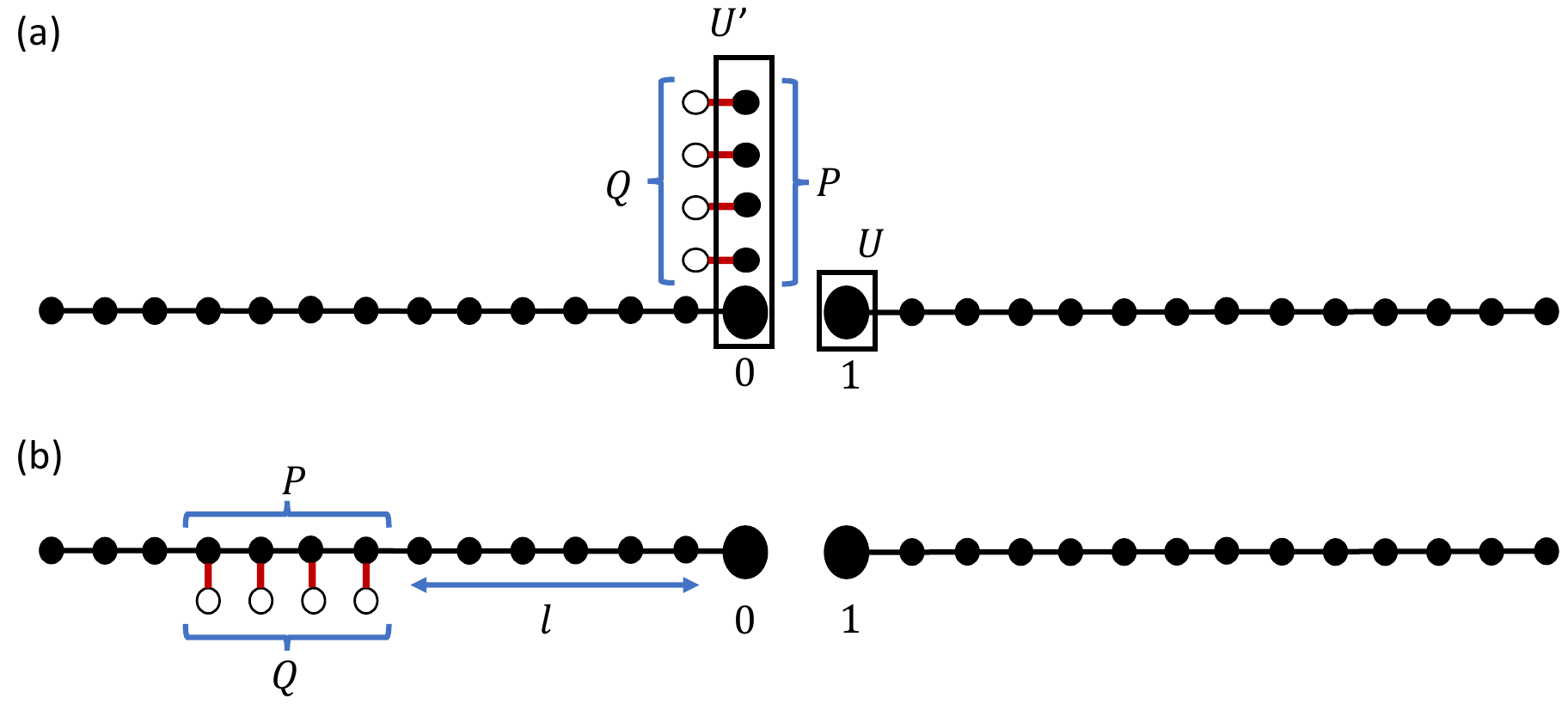}
\caption{Two different setups for studying information transfer between the black hole and the bath. In (a), a $p$-dimensional system $P$ is added to the black hole, and in the initial state $P$ is maximally entangled with a reference system $Q$. In (b), $p$ sites of the bath at a distance $l$ from the black hole site 0 are maximally entangled with a reference system $Q$ in the initial state, and the dynamics are the same as in Fig.~\ref{fig:bhm}. Red lines between pairs of sites indicate that they are maximally entangled in the initial state.}
\label{fig:ref_sys}
\end{figure}

Let us now explore how quantum information is transferred between an eternal black hole and its bath. 
We will consider two different processes: (i) the information was originally in the black hole; (ii) 
the information was originally outside the black hole, as shown in Fig.~\ref{fig:ref_sys}. We will see that the time scale $t_p$ and void formation again play 
a fundamental role. 

\subsubsection{Information was originally in the black hole} 

 Let us add to our setup an additional $p \ll k$ spins to site $0$, which we call {$P$}, within the ``left" black hole, and a $q^p$-dimensional reference system $Q$ which is maximally entangled with $P$. 
The dynamics of the ``left" black hole are modified so that we now have a time evolution operator $U'$ (which is again assumed to obey~\eqref{pgd}) acting on the union of site $0$ and $P$ at all even steps. The reference system $Q$ does not interact with any other system. 
We will take the initial state to be 
\be 
\rho_0 = \otimes_{i \leq -1} ~ \rho_i ~\otimes \rho_{01}~ \otimes \rho_{PQ}~\otimes_{i \geq 2}~ \rho_i 
\label{BH_initial_HP} 
\ee
where 
\be\label{kk}
\rho_{PQ} = \ket{\psi_{PQ}} \bra{\psi_{PQ}}, ~~~~ \ket{\psi_{PQ}}= \frac{1}{\sqrt{d_P}}\sum_{k=0}^{d_P-1} \ket{k}_P\ket{k}_Q , \quad d_P = q^p  \ .
\ee
We will now refer to the ``full'' black hole subsystem as $B = 0 \cup P \cup 1$. The bath $\sB$ consists of all the other lattice sites. 

As in our previous discussion for an evaporating black hole, $Q$ will remain maximally entangled with $B \cup \sB$ at all times, and 
\be \label{mne}
I_2 (Q, B; t) + I_2 (Q, \sB; t) = 2 S_2^{(Q)}(t)  = \sS_Q , \quad \sS_Q = \log d_P = p \log q  \ .
\ee
The calculation of $S_2^{(B)}(t)$ and $S_2^{(BQ)}(t)$ is very similar to our earlier discussion, and we find 
\be 
S_2^{(B)}(t) = \begin{cases}  \sS_Q +2 \seq t  & t< k \\
 \sS_Q + 2 S_{\rm BH} & t\geq  k
\end{cases} , \qquad S_2^{(BQ)}(t) = \begin{cases}  2 \seq t   & t< k+p \\
 2 \sS_Q + 2 S_{\rm BH}   & t\geq  k+p
\end{cases} 
  \ .
\label{BH}
\ee
The Renyi mutual information between $Q$ and $B$  is then given by 
\begin{equation} 
I_2 (Q, B; t) = \begin{cases} 
2 \sS_Q & t< k \\
2 \sS_Q  - 2 \seq (t-k) & k < t < k+p \\
0 & t\geq k+p 
\end{cases}  \ . \label{ib}
\end{equation}
So the mutual information between the reference system and the black hole starts decreasing after the Page time, and quickly goes to zero at a time scale which is proportional to the size of the reference system. 

Similarly we can find the mutual information of $Q$ with the bath $\sB$, 
\be 
S_2^{(\sB)}(t) = \begin{cases}  2 t \seq & t< k+p \\
 2 \sS_{\rm BH} + 2 \sS_Q  & t\geq  k+p 
 \end{cases}, \qquad S_2^{(\sB Q)}(t) = \begin{cases} \sS_Q +  2 t \seq & t< k \\
 \sS_Q +  2 S_{\rm BH} & t\geq  k
 \end{cases}
 \label{bath}
\ee
where the second lines of both expressions are results of void formation. One readily sees that the above expressions
give 
\begin{equation} 
I_2 (Q, \sB; t) = \begin{cases} 
0 & t< k \\
2 \seq (t-k) & k < t < k+p \\
2 \sS_Q  & t\geq k+p 
\end{cases} \label {isb}  
\end{equation}
which satisfies~\eqref{mne}.
In particular, without the contributions from void formation, one would find $I_2 (Q, \sB; t) =0$ at all times, and thus the information would be lost.

\subsubsection{The information was originally outside the black hole}

Now suppose we modify the black hole and bath setup to include a reference system $Q$ that is maximally entangled with 
$p$ spins outside the black hole, which we will again denote as $P$, as shown in figure \ref{fig:ref_sys} (b). We will take $l \ll k$.  Now  the initial state is 
\be 
\rho_0 = \otimes_{j <-l-p} ~ \rho_j  ~\otimes \rho_{PQ} ~ \otimes_{ -l-1< j <0} ~\rho_j ~\otimes \rho_{01} \otimes_{i \geq 2} \rho_i
\label{BH_initial_HP_out} 
\ee
where $\rho_{PQ}$ is given by~\eqref{kk}. 

Again the calculation is very similar to previous ones, so we will be brief, mostly listing the results. 
We have 
\begin{equation} 
S_2^{(\sB)}(t) = \begin{cases}  \sS_Q + 2t \seq & t< l  \\
\sS_Q + (t + l) \seq & l \leq t< l+p  \\
2 t \seq & l+p \leq t < k+{p \ov 2} \\
\sS_Q + 2 \sS_{\rm BH} & t\geq k+{p \ov 2}
\end{cases}, \qquad 
S_2^{(\sB Q)}(t) = \begin{cases}  2t \seq   & t< k \\
2 \sS_{\rm BH}   & t\geq k  
\end{cases}
\label{Sbath_out}
\end{equation}
which lead to 
\be 
I_2 (Q, \sB; t) = \begin{cases} 2\sS_Q  & t< l  \\
2\sS_Q + (l-t) \seq & l \leq t< l+p  \\
\sS_Q & l+p \leq t < k \\
\sS_Q + 2(t -k) \seq  & k \leq t \leq k+{p \ov 2} \\
2 \sS_Q  & t\geq k+p/2  
\end{cases} \ .
\label{Ii1}
\ee

We can also find the mutual information between the black hole subsystem $B$ and $Q$ \footnote{{Note that in the calculation of $S_2^{(B\cup Q)}$, we need to take into account the small probability of deviation from the sharp light-cone growth discussed in Appendix \ref{sec:light_cone}, due to the larger phase space of contributing initial operators from the region $P$ in the bath compared to other regions.}}
\be 
I_2 (Q, B; t) = \begin{cases} 0  & t< l  \\
(t-l ) \seq & l \leq t< l+p  \\
 \sS_Q & l+p \leq t < k   \\
\sS_Q  -2(t-k) \seq & k \leq t < k+p/2  \\ 
0 & t\geq k+p/2  
\end{cases} \ .
\label{Ii2} 
\ee
Equation~\eqref{Ii1}--\eqref{Ii2} again satisfy the unitarity constraint~\eqref{mne}. 
They show that due to the light-cone spreading {in the bath}, part of the information in $P$ ``falls'' into the black hole.  
Before the Page time, there is a long period where the information originally in $P$ is shared between the black hole and the bath, with each having one half. The information is transferred back to the bath again shortly after the Page time.  
Again without void formation, the part of the information which falls into the black hole will be lost.

\subsection{Free  bath} \label{sec:free} 

We now examine the situation where the evolution of bath is described by~\eqref{qp_u}.
Under evolution with such a $V$ {in a setup without a black hole}, an initial product state will remain a product state 
with no entanglement generated. If the initial state has short-range entanglement, then $V$ can propagate 
the short-range entanglement to long-distances~\cite{Liu:2019svk}. Thus, this may be considered 
a model for a free system. Various aspects of entanglement growth in this ``free propagation'' model were
discussed in detail in~\cite{Liu:2019svk}.

\subsubsection{Evolution of entanglement for the black hole and bath}

Let us again consider $S_2$ for $B$ and $\mathcal{B}$ shown in Fig.~\ref{fig:bhm} (a), starting from the initial state \eqref{BH_initial}. 
We find 
\be \label{ent_g_qp}
e^{- S_2^{(B)}(t) } = e^{- S_2^{(\mathcal{B})}(t) } = 
e^{- \seq t} + e^{- 2 \sS_{\rm BH}} \quad \Rightarrow \quad 
S_2^{(B)}(t) = S_2^{(\mathcal{B})}(t) = \begin{cases} 
t  \seq &  t <2k  \\
2\sS_{\rm BH}  & t\geq 2k 
\end{cases}  
\ee
{where we have taken $k \gg t\gg 1$.}
Like in the chaotic bath model, the growth of $S_2^{(B)}(t)$ and $S_2^{(\mathcal{B})}(t)$  is due to operator growth in the black hole, and the saturation of $S_2^{(\mathcal{B})}$ at the Page time $t_p = 2k$ results from void formation. Note that compared to the result \eqref{uenl} for random unitary circuits from the same initial state, the Page time is twice as long.

\begin{figure}[!h] 
 \centering
 \includegraphics[width=11cm]{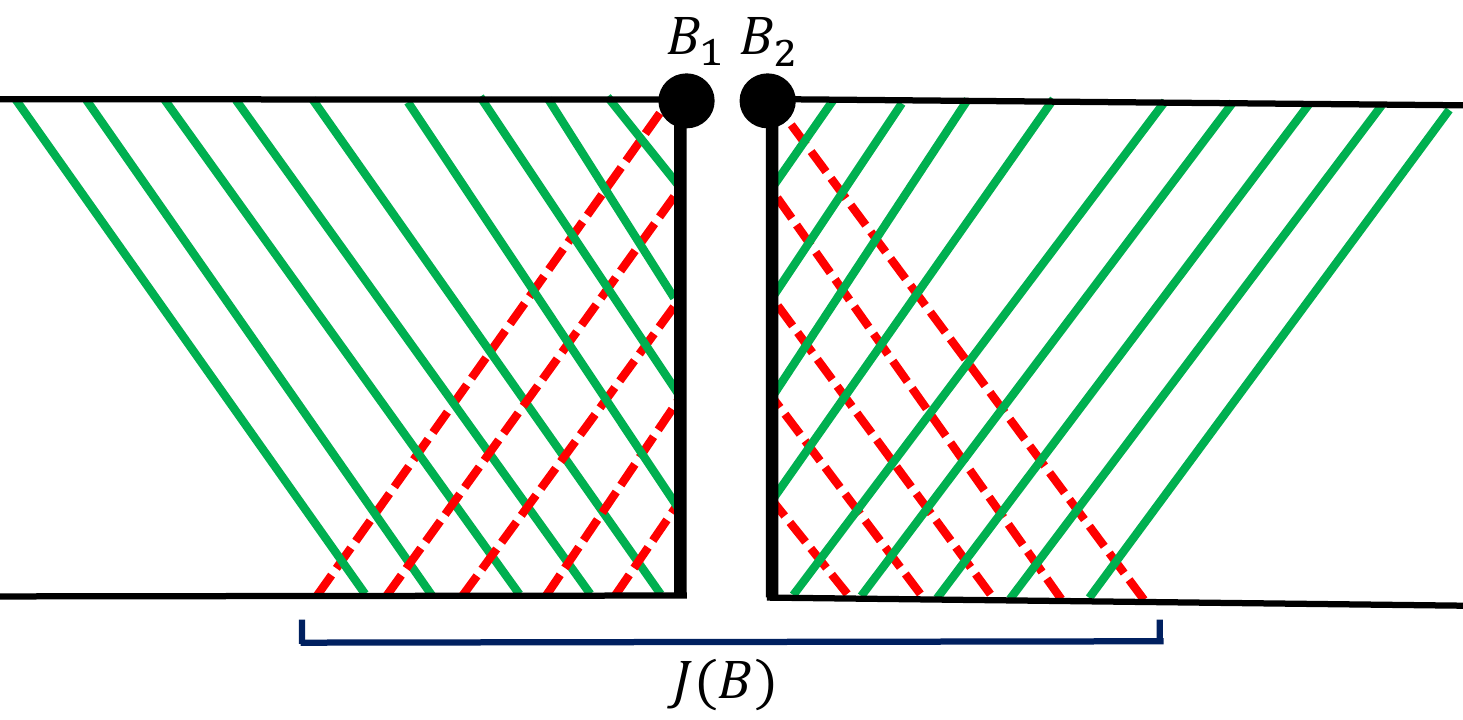}
 \caption{Operator growth in the toy model of a black hole with a free bath.}
 \label{fig:op_qp}
 \end{figure}

To see~\eqref{ent_g_qp}, first note that $V$ acting on sites $i, j$ simply translates operators from site $i$ to $j$ and vice versa, 
\be 
V^{\dagger} (O_i \otimes P_j) V = P_i \otimes O_j 
\ee
where $O, P$ are any single-site operators. As a result, operators
 at different sites evolve independently from each other, and operators at alternate sites move respectively to the left and right at speed 1. The resulting trajectories of initial operators from different sites in $J(B)$ are shown in Fig.~\ref{fig:op_qp}. Trajectories that take operators toward $B$ are represented by red dashed lines, while those that take operators away from $B$ are shown with solid green lines. It is clear that all initial operators outside $J(B)$, whose trajectories are not explicitly shown, will end up in $\sB$ at time $t$. 

Let us first find $N_B(t)$. It this model, every interaction of the black hole with the bath takes all operators non-trivial at sites $0_0$ and $1_0$ out of $B$ via the solid trajectories, and at the same time also brings all operators from two sites in $\sB$ into $B$ via the dashed trajectories. To estimate the factor by which $N_B(t)$ decreases due to the former process, we use reasoning similar to the derivation of $N_B(t)$ in the chaotic bath case. Between any two steps where the black hole interacts with the bath, a chaotic unitary evolution $U$ is applied within the black hole. Under the action of $U$, the probability that the final operator is trivial at both $0_0$ and $1_0$ is $q^{-4}$ from the random void distribution \eqref{pgd}. Thus, $N(t-2) q^{-4}$ operators out of the operators originally in $B$ at time $t-2$ remain in $B$ at time $t$, but in addition $q^2$ operators from two sites in $\sB$ are brought into $0_0$ and $1_0$ via the dashed trajectories, increasing $N(t)$ by a factor $q^2$. We therefore have 
\be 
N_B(t) = N_B(t-2) q^{-2}. 
\ee 
Since $N_B(t=0) = q^{2k}$, this implies $N_B(t) = q^{2k-t}$. Then using \eqref{hjn}, we obtain \eqref{ent_g_qp}. 

Note that if we considered an initial state in the bath consisting of entangled pairs between adjacent sites like in \cite{Liu:2019svk}, 
\be 
\sigma_0 = \otimes_{i<0}~ \tau_{i-1, i} ~ \otimes \rho_{01} ~\otimes_{i>1} ~\tau_{i, i+1}
\ee
where $\rho_{01}$ is as defined in \eqref{urn} and $\tau_{i,j}$ is as defined in \eqref{tau_ij}, then we would have sharp light-cone growth of all initial operators. In this case, there are  no operators from outside $B$ which can become localized in $B$ via the dashed trajectories (since each initial operator has one endpoint which propagates away from the black hole at all times). Hence, we would again have \eqref{nb_decrease}, and the entanglement growth would be given by \eqref{uenl}, with Page time $t_p = k$. 
 
 Now let us understand the evolution of $S_2$ for $\mathcal{B}$ with the initial state \eqref{BH_initial}. All initial operators contained in the complement of $J(B)$, as well as at the $t$ sites in $J(B)$ from which operators propagate to $\sB$ (the starting points of the solid green trajectories in $J(B)$ in figure \ref{fig:op_qp}), are localized in $B$ at time $t$. There are $q^{|\sB|-2t +t}$ such operators. At the remaining initial sites (including 0 and 1), we can either have the identity, in which case there is a probability 1 of being contained in $\sB$ at time $t$, or non-trivial operators, which can become contained in $\sB$ by forming a void in the complement of $1_0$ and $0_0$ in $B$ after $U$ is applied at the final step of the evolution. The probability of forming such a void is $q^{-2(2k-2)}\approx q^{-4k}$ from the random void distribution, and the number of contributing initial operators is $q^{2k+t}$, so we find 
 \be 
 N_{\sB}(t) =  q^{|\sB|-t} (1+ q^{2k+t} q^{-4k}) = q^{|\sB|-t} (1+ q^{t -2k}) \label{nsbt}
 \ee
 which gives \eqref{ent_g_qp}. 
Without taking void formation into account, we would again see unbounded growth of the entanglement entropy of the bath in this model due to the first term in \eqref{nsbt}. 
 
\subsubsection{Transfer of information} 

We consider the same setups as in Sec.~\ref{sec:trans}. 
Let us first understand how information that is initially inside the black hole comes out, taking the initial state \eqref{BH_initial_HP}. We find that the entropies $\sB$ and $\sB \cup Q$ grow as:
\be 
S_2^{(\sB)}(t)= \begin{cases} 
t \,\seq & t<2k +2p \\
2\sS_{\rm BH}+2 \sS_Q & t \geq 2k  +2p 
\end{cases} , \quad 
S_2^{(\sB\cup Q)}= \begin{cases} \sS_{Q}+t \, \seq & t<2k\\
2\sS_{\rm BH}+\sS_Q & t\geq 2k
\end{cases} 
\ee
Hence, the time-evolution of the mutual information between $\sB$ and $Q$ is given by 
\be 
I(\sB, Q; t) = \begin{cases} 
0 & t< 2k\\
 t \, \seq -2\sS_{\rm BH}  & 2k < t< 2k+2p \\
2 \sS_Q & t \geq 2k+2p
\end{cases} 
\label{qpi1}
\ee
We can similarly find that the mutual information $I(B, Q; t) = 2 \sS_Q - I_2 (Q, \sB; t)$ at all times.  

In the case where the information is initially outside the black hole, so that the initial state is \eqref{BH_initial_HP_out}, we find 
\begin{equation} 
\begin{gathered} 
S_2^{(\sB)}(t) = \begin{cases}  \sS_Q + t \, \seq & t< l  \\
\sS_Q + (t/2 + l/2) \seq & l \leq t< l+p  \\
\sS_Q/2+ t ~\seq & l+p \leq t < 2k  + p/2 \\
\sS_Q + 2 \sS_{\rm BH} & t\geq k+p  
\end{cases}, \\
S_2^{(\sB Q)}(t) = \begin{cases}  t\, \seq   & t< l \\
(3t/2 -l/2)\seq & l< t< l+p \\
\sS_Q/2+ t \, \seq   & l+p < t< 2k - p/2  \\
2 \sS_{\rm BH}   & t\geq 2k- p/2  
\end{cases}
\label{Sbath_out_qp}
\end{gathered}
\end{equation}
which lead to 
\be 
I_2 (Q, \sB; t) = \begin{cases} 2\sS_Q  & t< l  \\
2\sS_Q + (l-t) \seq & l \leq t< l+p  \\
\sS_Q & l+p \leq t < 2k-p/2 \\
3\sS_Q/2- 2 \sS_{\rm BH} + t  \, \seq &  2k-p/2 \leq t < 2k + p/2  \\
2 \sS_Q  & t\geq 2k+p/2  
\end{cases} \ .
\label{qpi2}
\ee
We again find that mutual information between the black hole subsystem $B$ and $Q$ is given by $2 \sS_Q - I_2 (Q, \sB; t)$ at all times. The qualitative nature of the results \eqref{qpi1} and \eqref{qpi2} is similar to \eqref{isb} and \eqref{Ii1}. The information again starts to come out of the black hole at the counterpart of the page time, $t_p = 2k$, and comes out at approximately half the rate we found in the random circuit bath model. In this free propagation model, the reason for the value $\sS_Q$ of the mutual information at intermediate times is immediately clear, as exactly half of the particles in $P$ propagate towards the black hole, while the rest propagate in the opposite direction. 

\section{Conclusions and discussion} \label{sec:conc}

In this paper, we developed an operator gas approach to studying 
simple models of evaporating as well as eternal black holes. 
We showed that the Page curve and the unitarity of evolution of entanglement 
are general consequences of void formation, and in particular of the random void distribution of chaotic systems. 
While the models we considered are rather crude, the results should {also apply to more realistic models of black holes}, as our discussion only requires broad aspects of these models {which should be present in any chaotic system}.  
This dynamical approach to deriving the Page curve also sidesteps the issue whether 
the state of a black hole and its radiation is ``typical,'' and hence potentially extends the 
validity of the Page curve to more general systems {than the ones that the original argument could be applied to}. 

Our results also resonate nicely with recent semi-classical gravity discussions of the Page curve for two-dimensional black holes, suggesting that void formation should underlie the semi-classical prescription of inclusion of ``islands'' and 
recent Euclidean replica wormhole calculations~\cite{pen, Al1, Al2, Al3, 
Rozali:2019day, 
Akers:2019nfi, 
Chen:2019uhq, 
Al4, 
Penington:2019kki, 
Almheiri:2019qdq}.  
 
In this paper we looked at the second Renyi entropy, {which has a simple relation to operator growth probabilities}, for technical simplicity. 
It would be nice to generalize the argument for higher Renyi and von Neumann entropies. 
Moreover, the models we considered are too simple to make direct connections with 
semi-classical gravity analysis. It would be interesting to extend our analysis to models 
such as SYK where there are closer connections to gravity, and where in principle it is possible to directly probe the operator growth probabilities. 

Our discussion also showed that void formation processes play a key role in the transfer of information from 
a black hole to its radiation, or to the bath surrounding it in the case of an eternal black hole. It would be interesting 
to explore whether one can use this insight to develop new algorithms for decoding the information in the radiation. 

Finally, as emphasized in~\cite{Liu:2019svk}, void formation processes are ubiquitous 
in quantum many-body systems in maintaining unitarity, and generating mutual information and multi-partite entanglement. 
If the connection with ``replica wormholes''  can be made more precise, this could imply that  replica wormholes 
 are not exotic objects, and likely are present in some form in calculations of the Renyi and von Neumann entropies for multiple subsystems in setups without a black hole.

\vspace{0.2in}   \centerline{\bf{Acknowledgements}} \vspace{0.2in}
We would like to thank Netta Engelhardt, Daniel Harlow, Aram Harrow, Sam Leutheusser, Lampros Lamprou, and Manuel Vielma 
 for discussions. 
This work is supported by the Office of High Energy Physics of U.S. Department of Energy under grant Contract Number  DE-SC0012567.

\appendix

\section{Examples of ``operator gases"}
\label{sec:op_app}

In this appendix, we derive the forms of the ``operator gas" associated with various initial density matrices that are used throughout the paper. 

Suppose have a system $L$ with $k$ sites, such that the Hilbert space at each site has dimension $q$, and is spanned by an orthonormal basis $\{ \ket{0}, ..., \ket{q-1}\}$. For a pure product state whose factors at all sites are the same, that is, a state of the form 
\be 
 \rho = \otimes_i \sigma_i , ~~~ \sigma= \ket{\phi} \bra{\phi} \label{product}
\ee
where $\ket{\phi}$ is some fixed state in the one-site Hilbert space, we can change our basis so that we can write 
$\sigma = \ket{0} \bra{0}$. 

We can introduce a basis of operators at each site,  
\be 
O_c = X^{s_1} Z^{s_2}, ~~~~ s_1, s_2 = 0, 1, ..., q-1 \label{obd_2}
\ee
where 
\be 
Z = \sum_{k=0}^{q-1}e^{2\pi i k/q}\ket{k}\bra{k}, ~~~~ X = \sum_{k=0}^{q-1}\ket{k+1}\bra{k} \ .
\ee
Note that an orthonormal basis of operators $\{\sO_{\alpha}\}$ for the entire system satisfying \eqref{hne}  can be obtained by constructing the tensor products $\otimes_i O_{c_i}$ for all possible sequences $c_i$ of numbers between 0 and $q^2-1$. 

In terms of the single-site basis \eqref{obd_2}, we can write
\be 
\sigma = \ket{0} \bra{0} = \frac{1}{q}\sum_{k=0}^{q-1} Z^k 
\ee
and as a result, $\rho$ can be written as
\be  \label{tyn}
\rho = \frac{1}{q^k}\sum_{a \in I} \sO_{a} 
\ee
where $a \in I$ corresponds to the requirement that all $\sO_{\alpha}$ appearing in the sum are of the form $\otimes_{i} Z^{s_i}$. 

Next, note that any pure state $\tau$ in the system $L$ (not necessarily a product state) can be obtained by a unitary transformation of $\rho$ defined in \eqref{product}, so that we have 
\be 
\tau = \frac{1}{q^k} \sum_{a \in I'} \sP_{a} 
\ee
where now $a \in I'$ corresponds to the requirement that all $\sP_{a}$ appearing in the sum are of the form $U(\otimes_{i} Z^{s_i})U^{\dagger}$, for some unitary $U$. Note that: 
\ben
\item Since all $\sO_{a}$ in~\eqref{tyn} satisfy \eqref{hne}, all $\sP_{a}$ also satisfy that condition among themselves. 
\item Since all $\sO_{a}$ in $I$ are mutually commuting, all $\sP_{a}$ in $I'$ are also mutually commuting. 
\item Since $\mathbf{1}_L$ is contained in $I$, it is also contained in $I'$. 
\een

Having seen how to write pure product states and arbitrary pure states in terms of  ``operator gases", let us now consider a maximally entangled state between two systems $L_1$ and $L_2$ of the same dimension $d$,
\be 
\ket{\psi} = {1 \ov \sqrt{d}} \sum_{i=0}^{q-1} \ket{i} \otimes \ket{\tilde i}
\ee
with the  density operator 
\be 
\rho = {1 \ov d} \sum_{i,j} \sA_{ij} \otimes \tilde \sA_{ij} , \qquad \sA_{ij} = \ket{i} \bra{j} , \quad   \tilde \sA_{ij} = \ket{\tilde i} \bra{\tilde j}
\ .
\ee
where $\ket{i}$ and $\ket{\tilde{i}}$ can be distinct bases in $L_1$ and $L_2$. 
Denoting $ij$ as $I$, we have 
\be 
\Tr (\sA_{I}^\da \sA_{J}) =  \de_{IJ}. \qquad \Tr (\tilde \sA_{I}^\da \tilde \sA_{J}) =  \de_{IJ}  \ .
 \ee

We can expand $\sA_I$ in terms of any basis $\sQ_\al$ satisfying $\Tr (\sQ_\al^\da \sQ_\b) = d \de_{\al \b}$ as  
\be
\sA_{I} = {1 \ov \sqrt{d}} \sum_\al c_{I \al} \sQ_\al   
\ee
with $c_{I \al}$ a unitary matrix. We can introduce $d^2$ operators $\tilde \sQ_\al$ which are related to $\tilde \sA_I$ by
\be
\tilde \sA_{I} = {1 \ov \sqrt{d}} \sum_\al c_{I \al}^* \tilde \sQ_\al  \ .
\ee
Since $\{c_{I \al}^*\}$ is also a unitary transformation, $\{\tilde \sQ_\al\}$ is an orthonormal basis normalized by 
\be 
 \Tr (\tilde \sQ_\al^\da \tilde \sQ_\b) = d \de_{\al \b} \ . 
\ee

We then find that 
\be 
\rho = {1 \ov d^2} \sum_I \sum_{\al , \b} c_{I \al} c_{I \b}^* \sQ_\al \otimes \tilde \sQ_\b = {1 \ov d^2} \sum_\al \sQ_\al \otimes 
\tilde \sQ_\al 
\ee 
where $\tilde \sQ_\al$ in any term in the sum is fixed when $\sQ_\al$ is fixed, and in particular when $\sQ_{\alpha}$ is the identity, $\tilde \sQ_{\alpha}$ is also the identity, and vice versa. 

\section{Sharp light cone growth from the random void distribution in the chaotic bath}
\label{sec:light_cone}

\begin{figure}[!h] 
\includegraphics[width=16cm]{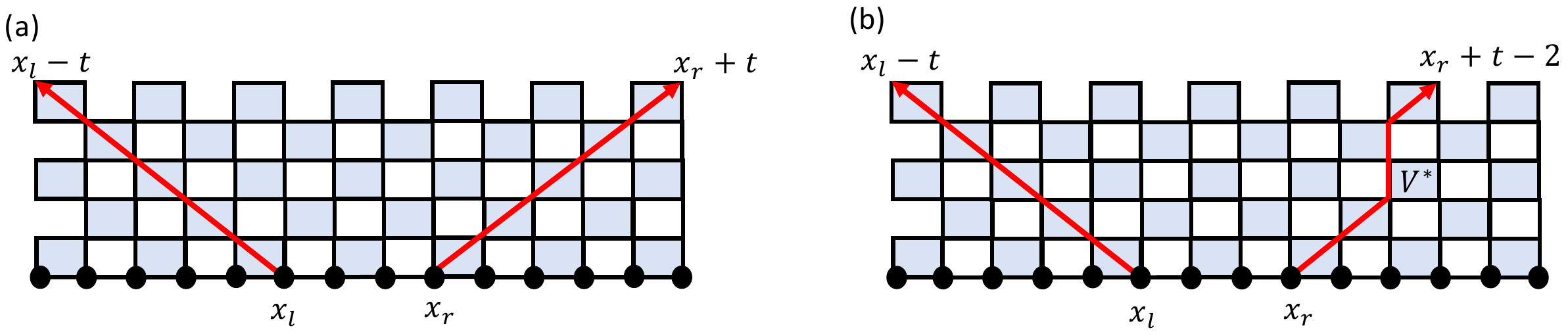}
\caption{Time-evolution of operators in the chaotic bath. All shaded rectangles represent unitary matrices $V$. (a) In a typical operator evolution process, no voids are formed under the action of any $V$ in the circuit, and we see sharp light-cone growth of the operator. (b) When a void is formed under the action of a single unitary matrix $V^*$ in the circuit, it shifts the right-endpoint of the final operator away from the edge of the light cone by two sites.}
\label{fig:voids_lc}
\end{figure}

Consider one of the unitary matrices $V$ applied in the chaotic bath, acting on sites $i, i+1$. We can assume that the scrambling time for the system $\{i,i+1\}$ is a single time-step, and hence can apply the random void distribution to each of the two sites $i$ and $i+1$ after the action of $V$, if $q$ is sufficiently large. We thus find that the probability of going from any non-trivial operator on $\{i,i+1\}$ to final operators trivial on any one site is $1/q^2$. The probability of going to operators non-trivial on both sites is thus close to $1$. Since this is the typical behaviour for each of the unitaries $V$, we find that an initial operator with endpoints $x_l, x_r$ evolves with probability approximately 1 to a final operator with endpoints $x_l-t$, $x_r+t$, as shown in figure \ref{fig:voids_lc}(a). Hence, the sharp light-cone growth in the chaotic bath can be seen as a consequence of assuming the random void distribution for each $V$. 

There is, however, a small non-zero probability of evolving to operators with endpoints $x_l-t+ \Delta_l$, $x_r+t -\Delta_r$, for $\Delta_l$ and $\Delta_r$ greater than zero. This requires the formation of voids under the action of $\Delta_l/2+  \Delta_r/2$ unitary matrices $V$ at intermediate steps, since the formation of a void after a single $V$ causes a deviation of the end-point from the edge of the light-cone by two sites, as shown in figure \ref{fig:voids_lc}(b). Since the probability of forming a void under the action of each $V$ is given by $1/q^2$, such a process has probability $q^{-\Delta_l -\Delta_r}$. 

This small probability can be neglected in all calculations of $S_2^{(A)}$ for various regions $A$ in the chaotic bath setup that we considered in Sec.~\ref{sec:ebh}, except in the case where $A=B\cup Q$, needed for finding the evolution of the mutual information of the black hole with a reference system initially entangled with a subsystem $P$ of the bath in \eqref{Ii2}. In that case, when initial operators from $P$ come into causal contact with the black hole, they give a contribution to $N_B(t)$ that cannot be neglected despite the smallness of the probability that they become localized in $B$, as each site from $P$ contributes $q^2$ rather than $q$ initial operators, increasing the ``phase space" factor in the contribution to $N_B(t)$.


\begin{thebibliography}{99}

\bibitem{Hawking}{S. Hawking, {\it Particle Creation by Black Holes, Commun.Math.Phys.} {\bf 43}  (1975)
199-220.}

\bibitem{Hawking2}{S. Hawking, {\it Breakdown of Predictability in Gravitational Collapse, Phys. Rev.} {\bf D14}  (1976)
 2460–2473.}

\bibitem{page} D. N. Page, ``Information in black hole radiation'', 
{\it Phys. Rev. Lett.} {\bf 71} (1993) 3743-3746, [arXiv:hep-th/9306083 [hep-th]].

\bibitem{page2}
D.~N.~Page, ``Average entropy of a subsystem," {\it Phys. Rev. Lett.} {\bf 71} (1993), 1291, [	arXiv:gr-qc/9305007].

\bibitem{lubkin}
E. Lubkin, {\it J. Math. Phys.} {\bf 19}, 1028 (1978)

\bibitem{Lloyd} 
S. Lloyd and H. Pagels, {\it Ann. Phys. (N.Y.)} {\bf 188}, 186 (1988).

\bibitem{susskind} 
Y. Sekino and L. Susskind, ``Fast scramblers,'' JHEP {\bf 10} (2008) 065,  arXiv:0808.2096  [hep-th]. 

\bibitem{butterfly}
S. Shenker and D. Stanford, ``Black holes and the butterfly effect." JHEP {\bf 03} (2014) 067,   arXiv:1306.0622 [hep-th].

\bibitem{Hayden:2007cs} 
  P.~Hayden and J.~Preskill,
  JHEP {\bf 0709}, 120 (2007)
  [arXiv:0708.4025 [hep-th]].

\bibitem{Liu:2019svk} 
  H.~Liu and S.~Vardhan,
  ``Void formation in operator growth, entanglement, and unitarity,''
  arXiv:1912.08918 [quant-ph].

\bibitem{pen} G. Penington, ``Entanglement Wedge Reconstruction and the Information Paradox,''  arXiv:1905.08255.

\bibitem{Al1}  A. Almheiri, N. Engelhardt, D. Marolf and H. Maxfield, ``The entropy of bulk quantum fields and the entanglement wedge of an evaporating black hole,'' arXiv:1905.08762.

\bibitem{Al2}  A. Almheiri, R. Mahajan, J. Maldacena and Y. Zhao, ``The Page curve of Hawking radiation from semiclassical geometry,'' arXiv:1908.10996.

\bibitem{Al3}  A. Almheiri, R. Mahajan and J. Maldacena, ``Islands outside the horizon,''  arXiv:1910.11077.

\bibitem{Rozali:2019day} 
  M.~Rozali, J.~Sully, M.~Van Raamsdonk, C.~Waddell and D.~Wakeham,
  arXiv:1910.12836 [hep-th].

\bibitem{Akers:2019nfi} 
  C.~Akers, N.~Engelhardt and D.~Harlow,
  arXiv:1910.00972 [hep-th].

\bibitem{Chen:2019uhq} 
  H.~Z.~Chen, Z.~Fisher, J.~Hernandez, R.~C.~Myers and S.~M.~Ruan,
  ``Information Flow in Black Hole Evaporation,''
  arXiv:1911.03402 [hep-th].

\bibitem{Al4}   A. Almheiri, R. Mahajan and J. E. Santos, ``Entanglement islands in higher dimensions,''  arXiv:1911.09666.

\bibitem{Penington:2019kki} 
  G.~Penington, S.~H.~Shenker, D.~Stanford and Z.~Yang,
  arXiv:1911.11977 [hep-th].

\bibitem{Almheiri:2019qdq} 
  A.~Almheiri, T.~Hartman, J.~Maldacena, E.~Shaghoulian and A.~Tajdini,
  arXiv:1911.12333 [hep-th].
  
\bibitem{netta}  N. Engelhardt and A. C. Wall, ``Quantum Extremal Surfaces: Holographic Entanglement
Entropy beyond the Classical Regime," JHEP {\bf 01}, 073 (2015), arXiv:1408.3203 [hep-th].


\bibitem{abanin}
W.W. Ho and D.A. Abanin, ``Entanglement Dynamics in Quantum Many-Body Systems,", {\it Phys. Rev. B.} {\bf 95}, 094302 (2017), arXiv:1508.03784 [cond-mat.stat-mech]. 
 
 \bibitem{MS} M. Mezei and D. Stanford, ``On entanglement spreading
in chaotic systems,'' JHEP {\bf 05} (2017) 065, arXiv: 1608.05101.
 
\bibitem{Mathur:2014dia} 
  S.~D.~Mathur,
  ``What is the dual of two entangled CFTs?''
  arXiv:1402.6378 [hep-th].



\end{thebibliography}
\end{document}